\documentclass[aps,superscriptaddress,showpacs,showkeys,nofootinbib,floatfix]{revtex4}
\usepackage{graphicx,epsfig,wrapfig,amssymb}
\newcommand{\be}{\begin{equation}}
\newcommand{\ee}{\end{equation}}
\newcommand{\ba}{\begin{eqnarray}}
\newcommand{\ea}{\end{eqnarray}}
\newcommand{\la}{\langle}
\newcommand{\ra}{\rangle}
\newcommand{\di}{ {\rm d} }
\begin{document}
%===================  TITLE, AUTHORS, AFFILIATIONS ===================
\newcommand*{\UConn}{Department of Physics,
  University of Connecticut, Storrs, CT 06269, USA}\affiliation{\UConn}
\newcommand*{\Bochum}{Institut f{\"u}r Theoretische Physik II,
  Ruhr-Universit{\"a}t Bochum, D-44780 Bochum, Germany}\affiliation{\Bochum}
\newcommand*{\Temple}{Department of Physics, Barton Hall, 
  Temple University, Philadelphia, PA 19122-6082, U.S.A.}\affiliation{\Temple}

\title{Intrinsic transverse parton momenta in deeply inelastic reactions}
\author{P.~Schweitzer}\affiliation{\UConn}
\author{T.~Teckentrup}\affiliation{\Bochum}
\author{A.~Metz}\affiliation{\Temple}

\date{February 2010}
%===================  PREPRINT NUMBER, JOURNAL =======================
%\preprint{}
%===================  ABSTRACT =======================================

\begin{abstract}
Intrinsic transverse parton momenta $p_T$ play an important role in the
understanding of azimuthal/spin asymmetries in semi-inclusive 
deep-inelastic scattering (SIDIS) and the Drell-Yan process (DY).  
We review and update what is presently known about $p_T$ 
%intrinsic transverse parton momenta 
from these processes.
In particular, we address the question to which extent data support
the popular Gauss model for the $p_T$-distributions.
We find that the Gauss model works very well, and observe that the
intrinsic transverse momenta in SIDIS and DY are compatible,
which is a support for the factorization approach.
As a byproduct we recover a simple but practical way of taking into 
account the energy dependence of $p_T$-distributions.
\end{abstract}
\pacs{13.88.+e, % Polarization in interactions and scattering
      13.85.Ni, % Inclusive production with identified hadrons
      13.60.-r, % Photon and charged-lepton interactions with hadrons
      13.85.Qk} % Hadron-induced inclusive production with identified leptons,
                % photons, or other nonhadronic particles (energy > 10 GeV)
\keywords{Semi-inclusive deep inelastic scattering, Drell Yan process,
      intrinsic transverse parton momenta}
\maketitle

\tableofcontents

\newpage
%=== SECTION 1: INTRODUCTION =========================================
\section{Introduction}
\label{Sec:Introduction}

Intrinsic transverse parton momenta in hadrons can be probed in deeply 
inelastic reactions, such as SIDIS and DY, when adequate transverse 
momenta in the final state are measured. Here ``transverse''  means 
with respect to the hard momentum flow in the process, e.g.\ in SIDIS 
transverse momenta of produced hadrons with respect to the virtual photon.
Transverse parton momenta are described in terms of transverse momentum
dependent parton distribution functions (TMDs) or analogously generalized
fragmentation functions 
\cite{Collins:1981uw,Collins:1981uk,
Collins:1999dz,Collins:2003fm,Collins:2007ph,Hautmann:2007uw,Collins:2008ht}.

Although the concept of transverse parton momenta dates back to early days of QCD 
\cite{Collins:1981uw,Collins:1981uk,
Georgi:1977tv,Cahn:1978se,Konig:1982uk,Efremov:1980kz,Collins:1984kg}
the field has received continuous interest from theory
\cite{Sivers:1989cc,Efremov:1992pe,Collins:1992kk,Collins:1993kq,
Kotzinian:1994dv,Mulders:1995dh,Boer:1997nt,Boer:1997mf,Boer:1999mm,Bacchetta:1999kz} 
and important steps in the understanding of TMDs within
QCD were taken only recently 
\cite{Brodsky:2002cx,Collins:2002kn,Belitsky:2002sm,Ji:2004wu,Collins:2004nx}.
In particular, factorization theorems have been formulated, which extend previous 
work \cite{Collins:1981uk} and ensure that the productions of hadrons in SIDIS 
or dileptons in DY with small transverse momenta compared to the hard scale
factorize in hard scattering parts and universal non-perturbative 
objects: TMDs, fragmentations functions, and soft factors 
\cite{Ji:2004wu,Collins:2004nx}.

The appealing perspective of the TMD approach is that it may explain 
single spin or azimuthal asymmetries in various reactions 
\cite{Bunce:1976yb,Adams:1991rw,Adams:2003fx},
especially in SIDIS 
\cite{Aubert:1983cz,Arneodo:1986cf,Osipenko,Mkrtchyan:2007sr,Airapetian:2009jy,
Adams:1993hs,Breitweg:2000qh,Chekanov:2006gt,Airapetian:2002mf,Ageev:2006da,
Avakian:2003pk,Airapetian:1999tv,Airapetian:2004tw,Avakian:2005ps,Alexakhin:2005iw,
Airapetian:2005jc,Airapetian:2008sk,Giordano:2009hi,Kafer:2008ud}, DY 
\cite{Ito:1981XX,Badier:1981ti,Palestini:1985zc,Falciano:1986wk,Guanziroli:1987rp,
Conway:1989fs,Zhu:2006gx,Zhu:2008sj},  
or hadron production in $e^*e^-$-annihilations 
\cite{Abe:2005zx,Seidl:2008xc,Vossen:2009xz}.
In phenomenological studies of such data 
\cite{D'Alesio:2004up,Collins:2005ie,Anselmino:2005nn,
Oganesian:1997jq,Gamberg:2003ey,
Barone:2005kt,Barone:2006ws,Gamberg:2007wm,Zhang:2008ez,Barone:2008tn,Barone:2009hw,
Efremov:2004tp,Collins:2005rq,Bianconi:2005yj,Anselmino:2009st,Kang:2009sm,
Gamberg:2005ip,Bianconi:2006hc,Sissakian:2008th,Sissakian:2010zz,Bacchetta:2010si,
Sissakian:2005vd,Sissakian:2005yp,Lu:2009ip}
one often works in lowest order QCD 
(``tree level'') and neglects higher order QCD corrections, which includes 
soft factors as well as higher order corrections in the hard scattering.
  In some sense, the information on soft factors can be
  thought of as ``reshuffled'' into an effective
  description of involved TMDs or fragmentation functions.

As in such studies typically many novel TMDs are involved 
\cite{Metz:2004je,Bacchetta:2004zf,Goeke:2005hb,Bacchetta:2006tn}, 
it is popular to assume the so-called Gauss model. For example, in the 
case of the unpolarized TMD or fragmentation function one assumes 
\ba\label{Eq:Gauss-f1}
    f^a_1(x,p_T) &=& f^a_1(x)\;
    \frac{\exp(-p_T^2/\la p_T^2\ra)}{\pi\la p_T^2\ra} \;,\\
    \label{Eq:Gauss-D1}
    D^a_1(z,K_T) &=& D_1^a(z)\,
    \frac{\exp(-K_T^2/\la K_T^2\ra)}{\pi\la K_T^2\ra}\;,
\ea
where $p_T=|\vec{p}_T|$, and the normalization is 
$\int\di^2p_Tf^a_1(x,p_T)=f^a_1(x)$ and similarly for $D_1^a$.
Strictly speaking the Gauss widths $\la p_T^2\ra$, $\la K_T^2\ra$ 
could be flavor dependent, and $x$- or $z$-dependent, i.e.,\ the 
Ans\"atze (\ref{Eq:Gauss-f1},~\ref{Eq:Gauss-D1}) in general do not imply 
a factorized $x$- or $z$- and transverse momentum dependence.

Although in this way the unknown $p_T$-dependence is reduced to ``one 
number,'' a Gauss width which ``characterizes the $p_T$-dependence'' 
of a  TMD, and although the Ansatz has not the correct large-$p_T$
asymptotics \cite{Bacchetta:2008xw}, it nevertheless proves to be useful 
in many processes sensitive to intrinsic transverse momenta
\cite{D'Alesio:2004up,Anselmino:2005nn,Collins:2005ie}.
In fact, it was shown in \cite{D'Alesio:2004up} that the 
Gauss model provides a useful approximation in many processes 
sensitive to intrinsic transverse momenta. 

The information on the Gauss model parameters in 
(\ref{Eq:Gauss-f1},~\ref{Eq:Gauss-D1}) used in 
many recent works was due to two independent studies of SIDIS data 
\cite{Anselmino:2005nn,Collins:2005ie}.
In \cite{Anselmino:2005nn} the EMC data \cite{Arneodo:1986cf} on 
the azimuthal asymmetry $A_{UU}^{\cos\phi}$ in unpolarized SIDIS were 
used to fix the parameters in (\ref{Eq:Gauss-f1},~\ref{Eq:Gauss-D1}). 
In \cite{Collins:2005ie} the widths % $\la p_T^2\ra$, $\la K_T^2\ra$ 
were determined using (uncorrected for acceptance effects) mean values 
from HERMES on average transverse momenta $\la P_{h\perp}\ra$ 
of produced hadrons.
The results were found
\ba\label{Eq:old-fit-pT2-KT2}
   \la p_T^2\ra = \cases{0.25\,{\rm GeV}^2 & in \cite{Anselmino:2005nn},\cr
                         0.33\,{\rm GeV}^2 & in \cite{Collins:2005ie},}
   \;\;\;\;
   \la K_T^2\ra = \cases{0.20\,{\rm GeV}^2 & in \cite{Anselmino:2005nn},\cr
                         0.16\,{\rm GeV}^2 & in \cite{Collins:2005ie},}
\ea
which are consistent within the unestimated uncertainties involved in the 
analyses. Both studies were probably the best one could do at that time,
but are not free of criticism. 
In \cite{Anselmino:2005nn} it was {\sl assumed} the observable
$A_{UU}^{\cos\phi}$ is due to the Cahn effect only, omitting 
effects of other TMDs.
In \cite{Collins:2005ie} it was {\sl assumed} the acceptance 
corrections were not large.
In both works the very applicability of the Ansatz 
(\ref{Eq:Gauss-f1},~\ref{Eq:Gauss-D1}) was {\sl presumed}.

Meanwhile new data emerged, which improve the situation and
allow to test the Gauss Ansatz in SIDIS more thoroughly than it was
possible in \cite{D'Alesio:2004up,Collins:2005ie,Anselmino:2005nn}.
One of the aims of this study is to provide such tests.
In Sect.~\ref{Sec-2:SIDIS} after convincing ourselves that the Gauss model 
passes the tests imposed by the new data, we will be in the position to update 
the information (\ref{Eq:old-fit-pT2-KT2}) on the Gauss model parameters. 
We shall also discuss the EMC 
data on azimuthal asymmetries in the light of the updated information,
and comment on the Cahn- and Boer-Mulders effect.

Another purpose of this work is to test the Gauss Ansatz 
in the Drell-Yan process, and to determine the Gauss width of $f_1^a$.
Also here we shall discuss azimuthal asymmetries in unpolarized DY,
and comment on the Cahn- and Boer-Mulders effect, see Sec.~\ref{Sec-3:DY}.

Finally, having established the applicability of the Gauss model in SIDIS 
and DY, we will address the question whether the descriptions of $p_T$-dependences 
in the two processes are compatible. This is what one expects on the basis of the
factorization approach and universality arguments, but a meaningful comparison
of the effects requires to carefully  take into consideration the different energies
typically probed in  SIDIS and DY, see Sec.~\ref{Sec-4:s-dependence}.
Our conclusions are contained in Sec.~\ref{Sec-5:conclusions}.

\newpage
%=== SECTION 2: SIDIS ================================================
\section{Intrinsic transverse momenta in SIDIS}
\label{Sec-2:SIDIS}

%------ BEGIN FIGURE 1: Kinematics of SIDIS --------------------------
        \begin{wrapfigure}[13]{RD}{8cm}
        \includegraphics[width=7cm]{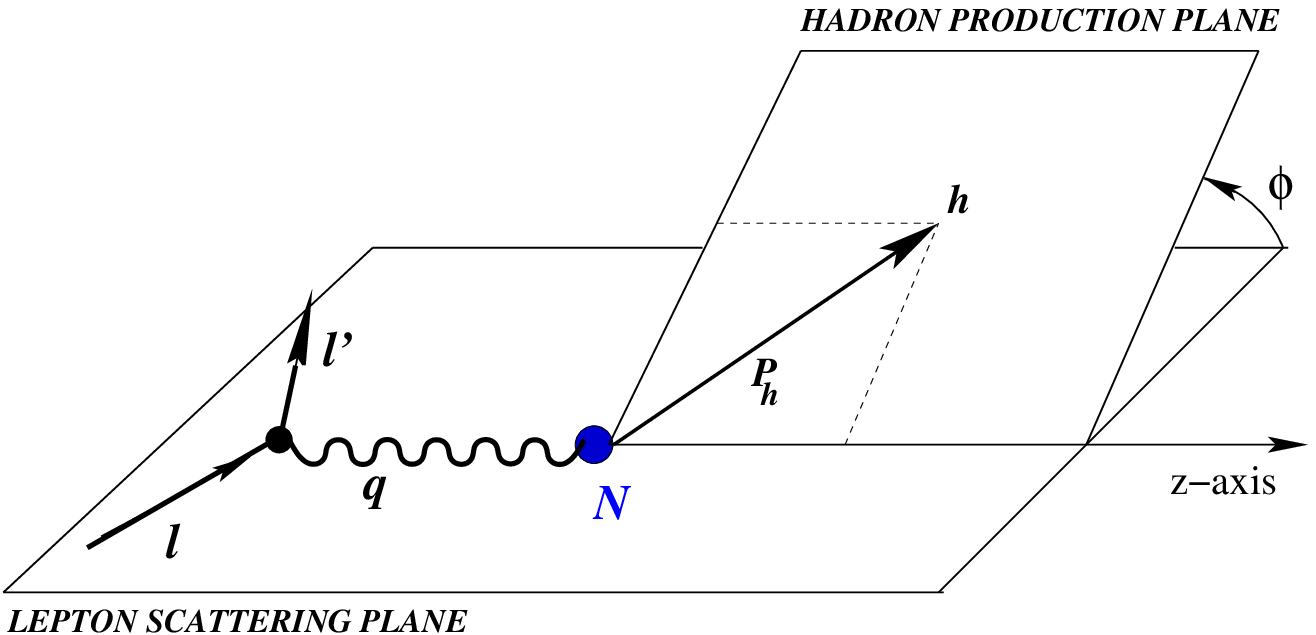}
        \caption{\label{Fig-01:kinematics}
        Kinematics of the SIDIS process $lp\to l^\prime h X$.}
        \end{wrapfigure}
%------ END FIGURE 1 -------------------------------------------------

In this Section we study the distributions of transverse hadron
momenta $P_{h\perp}$ in SIDIS. Two effects play a role, namely
intrinsic transverse parton momenta from the target which we model 
by (\ref{Eq:Gauss-f1}), and transverse momenta the hadrons acquire 
in the fragmentation process which we model by (\ref{Eq:Gauss-D1}).
After a brief introduction to SIDIS, we shall study in detail data 
from CLAS and HERMES, and comment then on the Cahn effect at EMC 
and forthcoming experiments.

Let $P$, $l$ ($l^\prime$) and $P_h$ denote (respectively) the momenta
of the proton, incoming (outgoing) lepton and produced hadron.
The relevant kinematic variables are $q= l-l'$ with $Q^2=- q^2$, 
$W^2=(P+q)^2$, $x=Q^2/(2P\cdot q)$, $y=(P\cdot q)/(P\cdot l)$, 
and $z=(P\cdot P_h)/(P\cdot q)$.

In the following we are interested in the transverse momentum $P_{h\perp}$ 
of the produced hadron, which is defined with respect to the momentum
of the virtual photon, see Fig.~\ref{Fig-01:kinematics}.
Assuming the Gauss model (\ref{Eq:Gauss-f1},~\ref{Eq:Gauss-D1})
the cross section differential in $x$, $y$, $z$, $P_{h\perp}^2$ 
(but averaged over the azimuthal angle $\phi$ of the produced hadron)
reads
\be\label{Eq:sigma0}
     \frac{\di^4\sigma_{UU}(x,y,z,P_{h\perp})}{
       \di x\,\di y\,\di z\,\di P_{h\perp}^2}=
     \frac{4\pi^2\alpha^2\,s}{Q^4}\;\biggl(1-y+\frac12y^2\biggr)\;
     \sum\limits_a e_a^2 \,x\,f_1^a(x)D_1^a(z)\;{\cal G}(P_{h\perp})
\ee
where the subscript ``$UU$'' indicates that both leptons and nucleons
are unpolarized. The function ${\cal G}(P_{h\perp})$ is given by
\be\label{Eq:def-G-kappa}
     {\cal G}(P_{h\perp})=\frac{1}{\pi\kappa_T^2(z)}\,
     \exp\biggl(-\;\frac{P_{h\perp}^2}{\kappa^2_T(z)}\;\biggr)\;,\;\;\;
     \kappa_T^2(z) = z^2\la p_T^2\ra+ \la K_T^2\ra\;,
\ee
with the normalization $\int\di^2P_{h\perp}{\cal G}(P_{h\perp})=1$, 
while the cross section (\ref{Eq:sigma0}) is normalized such that 
\be\label{Eq:sigma0-II}
     \frac{\di^3\sigma_{UU}(x,y,z)}{\di x\,\di y\,\di z} =
     \frac{4\pi\,\alpha^2\,s}{Q^4}\;\biggl(1-y+\frac12y^2\biggr)\;
     \sum\limits_a e_a^2 \,x\,f_1^a(x)D_1^a(z)
\ee
In the cross sections (\ref{Eq:sigma0}) and (\ref{Eq:sigma0-II}) 
we neglect power suppressed terms of the order ${\cal O}(M^2/Q^2)$ 
where $M$ is the nucleon mass. The neglected terms include purely 
kinematic factors, as well as a structure function ($F_{UU,L}$ 
in the notation of \cite{Bacchetta:2006tn}) which has no partonic 
description. Depending on the kinematics these contributions of 
${\cal O}(M^2/Q^2)$ need not to be small.
We will recall this when necessary, but we will not need their 
explicit forms here.

For later convenience we introduce also the notion of average 
transverse hadron momenta and their squares which are defined,
and in the Gauss model given, as 
\ba\label{Eq:Phperp-av}
        \la P_{h\perp}(z)\ra = 
        \biggl\la\frac{P_{h\perp}\,
         \di\sigma_{UU}(z,P_{h\perp})/\di z\,\di P_{h\perp}^2}
        {\di\sigma_{UU}(z,P_{h\perp})/\di z\,\di P_{h\perp}^2}\biggr\ra 
         &\stackrel{\rm Gauss}{=}& 
         \frac{\sqrt{\pi}}{2}\,\sqrt{\kappa_T^2(z)}\;,\\
\label{Eq:Phperp2-av}
        \la P_{h\perp}^2(z)\ra = 
        \biggl\la\frac{P_{h\perp}^2\,
         \di\sigma_{UU}(z,P_{h\perp})/\di z\,\di P_{h\perp}^2}
        {\di\sigma_{UU}(z,P_{h\perp})/\di z\,\di P_{h\perp}^2}\biggr\ra 
        &\stackrel{\rm Gauss}{=}&  \kappa_T^2(z)\;,
\ea
where $\la\;\dots\;\ra$ denotes average over $P_{h\perp}$. Analogously 
one could define mean transverse momenta as functions of other 
kinematic variables, for example $\la P_{h\perp}(x)\ra$.
Strictly speaking, in obtaining the Gauss model results in 
(\ref{Eq:Phperp-av},~\ref{Eq:Phperp2-av}) we assume the Gauss widths
to be flavor- and $x$-independent. As we shall see below,
these are reasonable approximations.

\newpage
%=== SECTION 2A: CLASS AND HALL C ====================================
\subsection{Lessons from CLAS and Hall-C}
\label{Subsec-2A:Lesson-from-JLab}

\begin{figure}[b!]
\begin{center}
\includegraphics[height=6cm]{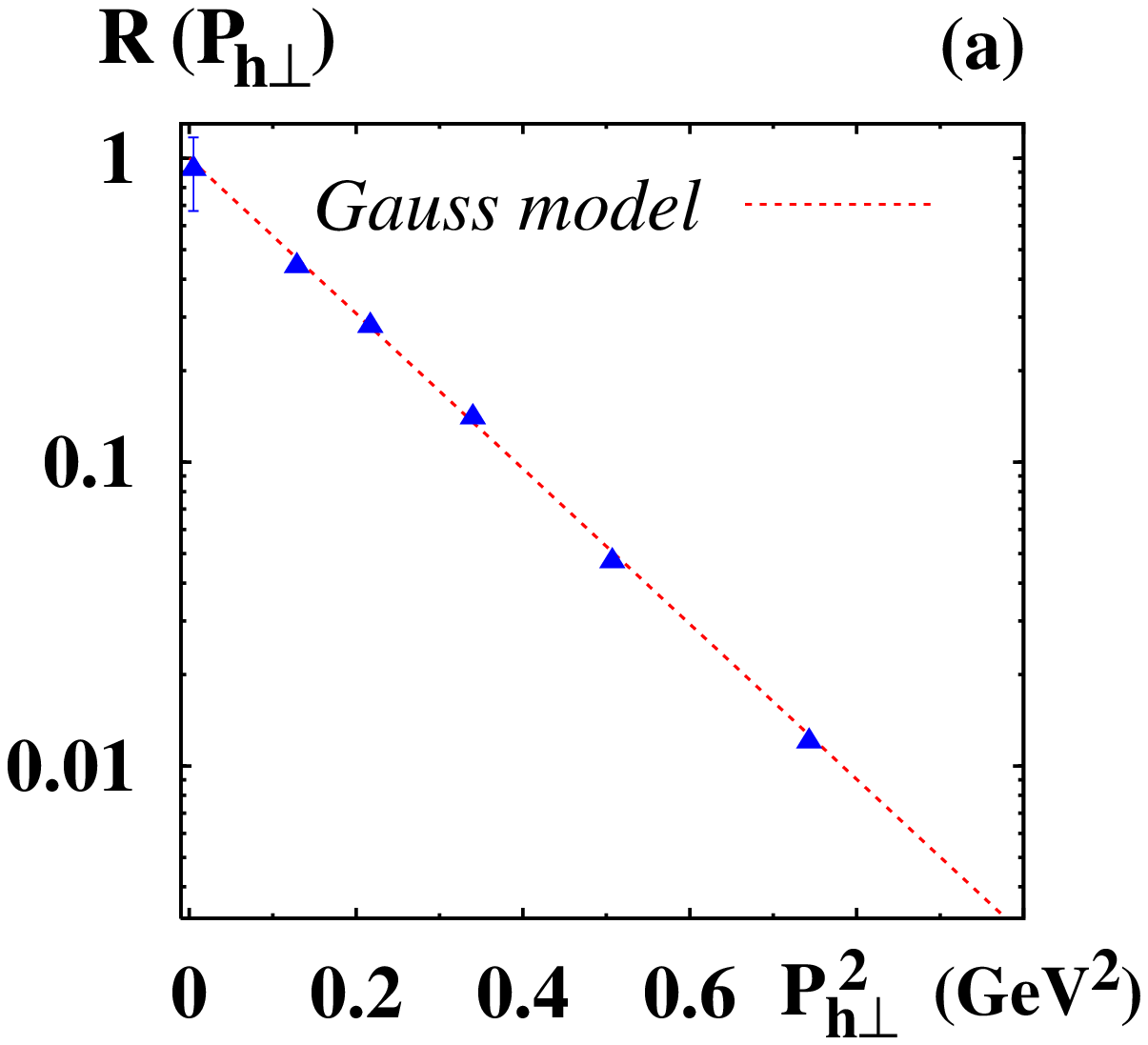}
\includegraphics[height=6cm]{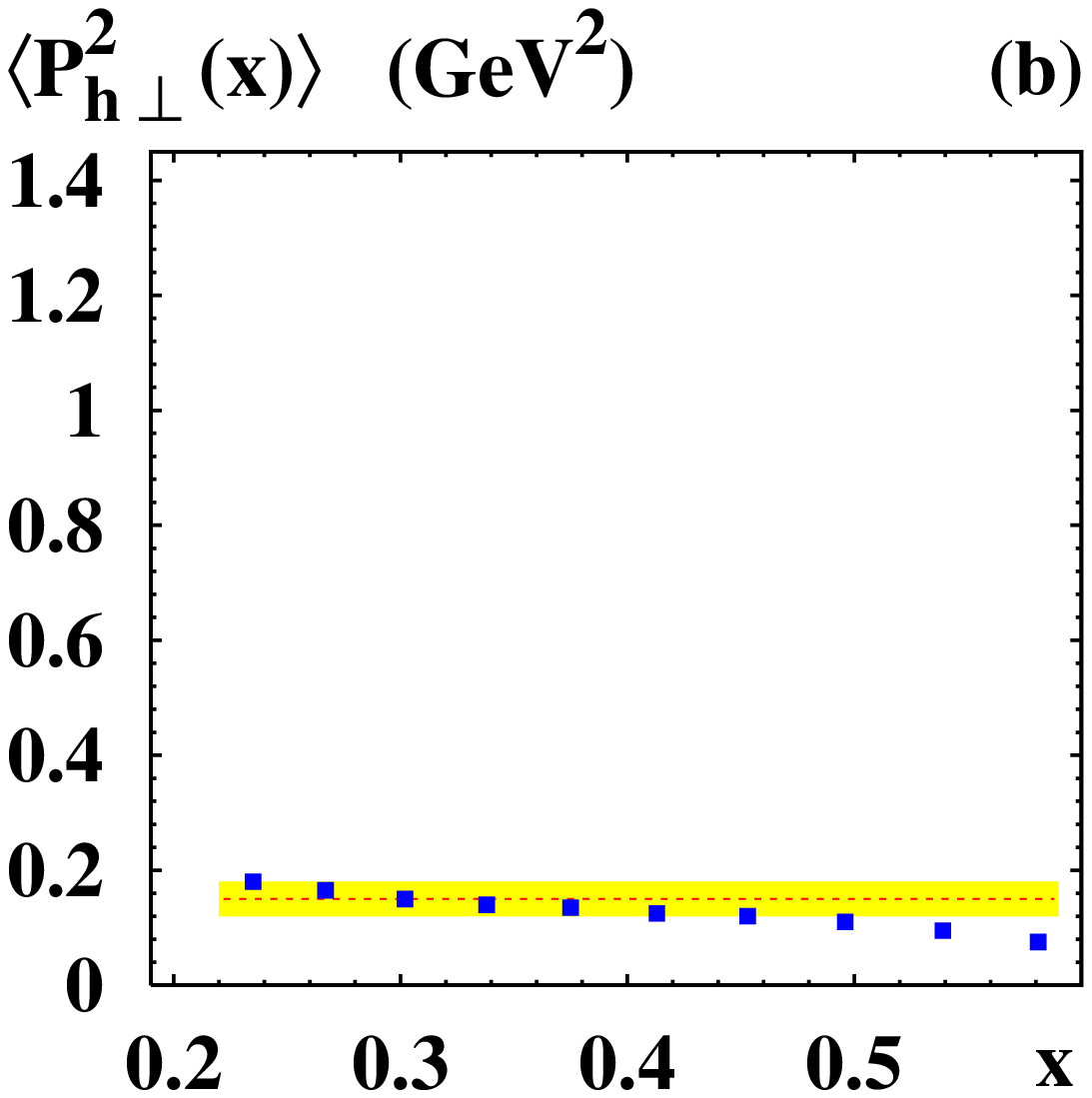}
\end{center}
        \vspace{-0.3cm}
        \caption{\label{Fig-02:CLAS}
	(a) The ratio $R(P_{h\perp})$ as defined in Eq.~(\ref{Eq:R})
        as function of the hadron transverse momentum square $P_{h\perp}^2$.
        The data are for $\pi^+$ from 
        CLAS \cite{Osipenko}. The dotted line is an effective
        description in the Gauss model, see text.
        (b) The average transverse momentum square $\la P_{h\perp}^2\ra$
        of $\pi^+$ produced at $z=0.34$ and $Q^2=2.37\,{\rm GeV}^2$ 
        in SIDIS at CLAS \cite{Osipenko} as function of $x$. 
        The dotted line is an effective
        description in the Gauss model assuming the Gauss width of
        $f_1^a(x,p_T)$ to be $x$-independent. This describes data
        within $20\%$ in the region $0.2<x<0.5$ as the shaded region
        shows.}
\end{figure}

In the CLAS experiment \cite{Osipenko} the semi-inclusive $\pi^+$ 
electro-production off a proton target was studied with a $5.75\,{\rm GeV}$ beam. 
Among others the following quantity was measured:
\be\label{Eq:R}
R(P_{h\perp})\;\equiv\;
\frac{\di^4\sigma_{UU}(x,y,z,P_{h\perp})/\di x\,\di y\,\di z\,\di P_{h\perp}^2}
     {\di^4\sigma_{UU}(x,y,z,0)/\di x\,\di y\,\di z\,\di P_{h\perp}^2}
     \;=\; \exp\biggl(-\;\frac{P_{h\perp}^2}{\kappa^2_T(z)}\biggr)\,,
\ee
at $x=0.24$ and $z=0.30$ for three different values of $Q^2$.
In the last step of (\ref{Eq:R}) we  assumed flavor-independent 
Gauss widths. For all three values of $Q^2$  the data are remarkably 
well described by the Gauss model \cite{Osipenko}.
In Fig.~\ref{Fig-02:CLAS}a we show the data on the ratio (\ref{Eq:R}) 
for the highest $Q^2=2.37\,{\rm GeV}^2$, which is very well described
with the parameter
\be\label{Eq:fit-CLAS-I}
     \kappa_T^2(z)\biggl|_{z=0.30} \;=\; 0.17\,{\rm GeV^2}\;.
\ee
The agreement of the data with the Gauss Ansatz is astonishing.
At lower $Q^2=1.74\,{\rm GeV}^2$ and $2\,{\rm GeV}^2$
the situation is equally impressive, with somewhat 
lower values for $\kappa_T^2(z)$ \cite{Osipenko}.

However, several reservations need to be made. 
First, in the CLAS kinematics \cite{Osipenko} 
the contributions of ${\cal O}(M^2/Q^2)$ we mentioned 
in the context of Eqs.~(\ref{Eq:sigma0}) and (\ref{Eq:sigma0-II})
are not negligible, i.e.,\ the description of CLAS data 
in Eqs.~(\ref{Eq:R},~\ref{Eq:fit-CLAS-I}) effectively 
parametrizes also these contributions.
Second, at $z=0.30$ the measured hadrons are not only due to
the fragmentation of the struck quark (``current fragmentation'')
but can also originate from the hadronization of the 
target remnant (``target fragmentation''). The latter is described
in terms of so-called fracture functions, which --- while being a 
fascinating topic by themselves --- are an undesired contamination
from the point of view of DIS.

%------ BEGIN FIGURE 2: SIDIS AT CLAS --------------------------------- 
%-------- END FIGURE 2. ----------------------------------------------- 

Nevertheless, although one has to keep in mind these reservations, 
presently the $5.75\,{\rm GeV}$ beam CLAS data \cite{Osipenko} provide 
the best support for the applicability of the Gauss model in SIDIS. 
It would be desirable to solidify this observation with data taken
at higher energies at CLAS12, HERMES and COMPASS.

Another conclusion one can draw from the CLAS data \cite{Osipenko},
modulo the above-mentioned reservations, is that the Gauss width 
$\la p_T^2\ra$ of $f_1^a(x,p_T)$ is only moderately $x$-dependent, 
see Fig.~\ref{Fig-02:CLAS}b which shows the average transverse momentum 
square $\la P_{h\perp}^2\ra$ of $\pi^+$ produced at $z=0.34$ and 
$Q^2=2.37\,{\rm GeV}^2$ at CLAS \cite{Osipenko} as function of $x$.
In fact, for $0.2<x<0.5$ we find 
$\la P_{h\perp}^2\ra = 0.15\,{\rm GeV}^2$ within $20\%$, which is 
demonstrated by the shaded region in Fig.~\ref{Fig-02:CLAS}b.

In the Gauss model, the $z$-dependence of the average hadron transverse 
momentum square $\la P_{h\perp}^2(z)\ra$ in Eq.~(\ref{Eq:Phperp2-av})
allows in principle to fix the Gauss widths $\la K_T^2\ra$ and 
$\la p_T^2\ra$ of $D_1^a$ and $f_1^a$. Such data were presented in 
\cite{Osipenko},
but here we will not use them for a quantitative determination of
the Gauss model parameters. In fact, at the moderate beam energies
for $z\lesssim 0.4$ \cite{Avakian:2003pk} the produced hadrons 
receive also contributions from target fragmentation, see above, and 
at still lower $z$ ``threshold effects'' play a role \cite{Osipenko}.
Therefore we refrain here from using these data quantitatively,
and will come back to them later for a qualitative comparison in 
Sec.~\ref{Subsec-2C:HERMES-fixing}.

Next we turn our attention to the Jefferson Lab data 
from the Hall-C collaboration \cite{Mkrtchyan:2007sr} 
where $5.5\,{\rm GeV}$ electrons were scattered off
proton and deuterium targets in the kinematics
$0.2 < x < 0.5$, $2\,{\rm GeV}^2 < Q^2 < 4 \,{\rm GeV}^2$,
$0.3 < z < 1$, and $\pi^\pm$ with transverse momenta up 
to $P_{h\perp}^2 < 0.2 \,{\rm GeV}^2$ were measured. 
In spite of the narrow $P_{h\perp}^2$-range covered,
the results on the differential cross-section
($\Omega$ and $E$ denote solid angle and energy of 
produced the hadron $h$ with $h=\pi^\pm$)
\be\label{Eq:sigma-Hall-C}
	\frac{\di^5\sigma(P_{h\perp}^2)_{et\to e^\prime hX}}
	{\di\Omega\,\di E\,\di z\,\di P_{h\perp}^2}
	\;\;\stackrel{\rm Gauss}{=}\;\;
	\frac{\di^5\sigma(0)_{et\to  e^\prime hX}}
	{\di\Omega\,\di E\,\di z\,\di P_{h\perp}^2}
	\;\exp\biggl(-\frac{P_{h\perp}^2}{\kappa_T^2(z)}\;\biggr)
\ee
allow a valuable cross-check at $\la x\ra=0.32$ and $\la z\ra=0.55$
($t=p,\,d$ denotes the proton, deuteron target).

%------ BEGIN FIGURE 3: SIDIS AT HALL-C ------------------------------- 
\begin{figure}[b!]
\begin{center}
        \vspace{-0.7cm}
\includegraphics[height=13cm]{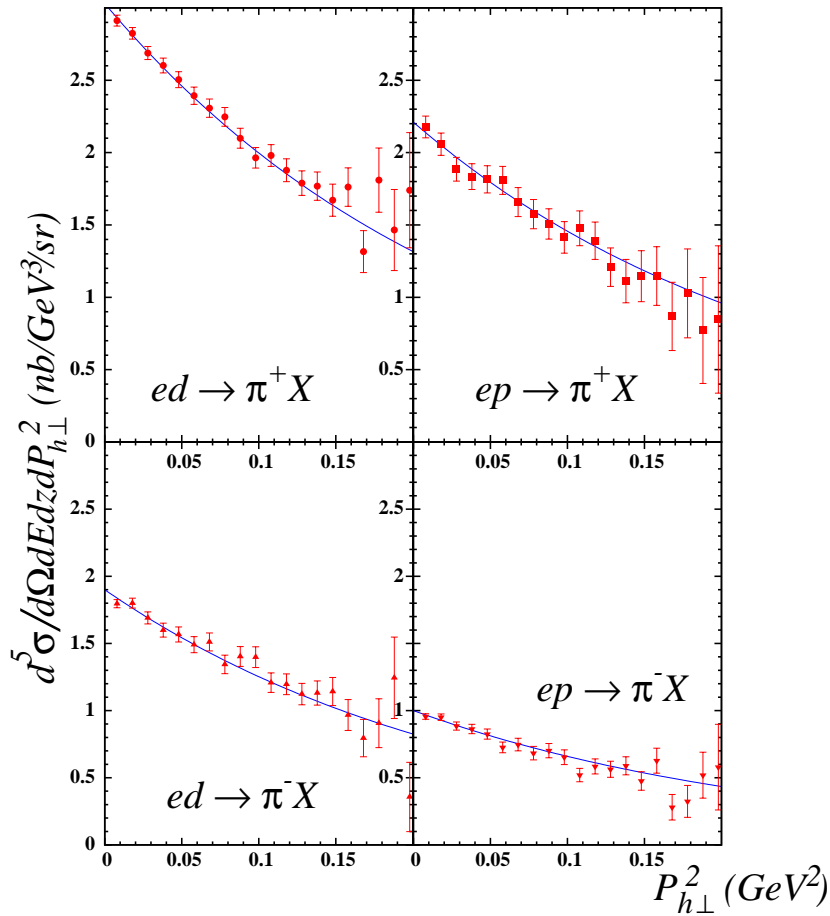}
\end{center}
        \vspace{-0.4cm}
        \caption{\label{Fig-03:Hall-C}
	The differential cross-section
	$\di^5\sigma/\di\Omega\,\di E\,\di z\,\di P_{h\perp}^2$
	for pion production off proton and deuterium targets
	at $\la x\ra=0.32$ and $\la z\ra=0.55$ as function 
	of $P_{h\perp}^2$ from Hall-C \cite{Mkrtchyan:2007sr}.
	The theoretical curves are from the Gauss model
	with the Gauss width fixed from CLAS \cite{Osipenko}.
	The overall normalization of the cross sections is fixed by hand.}
\end{figure}
%-------- END FIGURE 3. ----------------------------------------------- 

As the kinematics is similar to CLAS, we expect the Hall-C data,
which refer to $\la z\ra=0.55$, to be described in the Gauss model 
by the parameter $\kappa_T^2(z)=0.24\,{\rm GeV}^2$. 
This value is taken from CLAS data on $\la P_{h\perp}^2(z)\ra$
\cite{Osipenko} 
(see below, Fig.~\ref{Fig-04:HERMES-fixing-CLAS-cross-check}b).
In this way we obtain a very good description of the 
$P_{h\perp}^2$-dependence of the Hall-C data, see Fig.~\ref{Fig-03:Hall-C}. 

At this point it is interesting to stress that we 
assumed $\kappa_T^2(z)=0.24\,{\rm GeV}^2$ to be the same
for any hadron $h$ from any target $t$. This means, we 
assumed flavor-independent Gauss widths $\la p_T^2\ra$ 
and $\la K_T^2\ra$.
Thus, although a dedicated analysis in \cite{Mkrtchyan:2007sr} 
indicated a preference for slightly flavor-dependent Gauss 
widths, we conclude from Fig.~\ref{Fig-03:Hall-C} 
that the assumption of flavor-independent Gauss widths $\la p_T^2\ra$ 
and $\la K_T^2\ra$ is reasonable --- in the valence-$x$ region 
for $\la z\ra=0.55$.

To summarize, the data from Jefferson Lab \cite{Osipenko,Mkrtchyan:2007sr} 
are very well described assuming a Gauss distribution of intrinsic 
transverse parton momenta in the unpolarized distribution and fragmentation
functions, and suggest a weak flavor- and $x$-dependence of the Gauss 
width $\la p_T^2\ra$ of $f_1^a(x,p_T)$,
and a weak flavor-dependence of the Gauss widths $\la K_T^2\ra$ 
of $D_1^a(z,K_T)$. 
However, one has to bear in mind that these data contain contributions 
from fracture functions and/or terms of ${\cal O}(M^2/Q^2)$, such 
that we shall limit ourselves here to these qualitative conclusions, and 
use for a quantitative analysis data from HERMES \cite{Airapetian:2009jy},
see Secs.~\ref{Subsec-2B:HERMES-test-Gauss} and 
\ref{Subsec-2C:HERMES-fixing}.

\newpage
%=== SECTION 2B: PARAMETERS FROM HERMES ==============================
\subsection{Testing the Gauss model at HERMES}
\label{Subsec-2B:HERMES-test-Gauss}

%------ BEGIN FIGURE 4: Phperp(z) AND Phperp^2(z) at HERMES ----------
        \begin{wrapfigure}[20]{RD!}{6cm}
        \vspace{-0.6cm}
        \centering
        \includegraphics[width=2.2in]{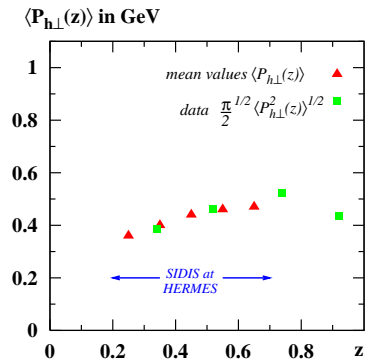}
        \vspace{-0.6cm}
        \caption{\label{Fig3-test-Gauss-HERMES}
        Transverse momenta of hadrons in SIDIS off deuterium 
	at HERMES vs.~$z$. We compare $\la P_{h\perp}(z)\ra$
        (triangles) from \cite{Airapetian:2002mf}, with 
        $\frac12\,\sqrt{\pi}\,\la P^2_{h\perp}(z)\ra^{1/2}$ (squares) from 
        \cite{Airapetian:2009jy}.
        In the indicated SIDIS range of HERMES 
        these quantities are predicted to coincide
        in the Gauss model, see Eq.~(\ref{Eq:Gauss-predict}).}
        \end{wrapfigure}
%------ END FIGURE 4 -------------------------------------------------

Assuming the Gauss model (\ref{Eq:Gauss-f1},~\ref{Eq:Gauss-D1}) 
and flavor- and $x$-independent Gauss widths, we obtained 
for the mean transverse hadron momenta and their squares the
results quoted in (\ref{Eq:Phperp-av},~\ref{Eq:Phperp2-av}). 
In particular, these two quantities are related in the Gauss 
model according to
\be\label{Eq:Gauss-predict}
       \la P_{h\perp}(z)\ra^2 = \frac{\pi}{4}\,\la P_{h\perp}^2(z)\ra\;.
\ee

Meanwhile data from HERMES on SIDIS of electrons or positrons with 
$E_{\rm beam}=27.6\,{\rm GeV}$ off a deuteron target allow to test the 
Gauss model prediction (\ref{Eq:Gauss-predict}) in the kinematics 
$Q^2>1\,{\rm GeV}^2$, $W^2>10\,{\rm GeV}^2$, $y<0.85$, $z>0.2$
and $0.023<x<0.4$ with $\la Q^2\ra = 2.4\,{\rm GeV}^2$, 
$\la x\ra = 0.09$ \cite{Airapetian:2002mf,Airapetian:2009jy}.

In Fig.~\ref{Fig3-test-Gauss-HERMES} we show $\la P_{h\perp}(z)\ra$ 
as function of $z$ from \cite{Airapetian:2002mf} (triangles). 
These mean values refer to pions and kaons and were not corrected 
for acceptance effects. We compare them with
$\frac12\sqrt{\pi}\la P^2_{h\perp}(z)\ra^{1/2}$ using the data from
\cite{Airapetian:2009jy} on $\la P^2_{h\perp}(z)\ra$ for positive pions.
At HERMES the SIDIS events are subject to the cuts $0.2 < z < 0.7$, 
as indicated in Fig.~\ref{Fig3-test-Gauss-HERMES}, 
and in this range $\pi^\pm$ and $K^+$ have very similar 
$\la P^2_{h\perp}(z)\ra$ \cite{Airapetian:2009jy}. 
We shall come back to this observation in the next section.
Let us add that the actual aim of  \cite{Airapetian:2009jy} was 
to study nuclear $P_{h\perp}$-broadening effects, the so-called 
Cronin-effect, and various nuclear targets were used besides 
deuteron, namely He, Ne, Kr, Xe.

We conclude from the exercise in Fig.~\ref{Fig3-test-Gauss-HERMES}
that the SIDIS data from HERMES support very well the Gauss 
model relation (\ref{Eq:Gauss-predict}), even though one should keep 
in mind possible effects due to acceptance corrections in the case
of the  $\la P_{h\perp}(z)\ra$ results from \cite{Airapetian:2002mf}.
Of course, data on the cross section differential in $P_{h\perp}$ from HERMES, 
of the type as presented by CLAS \cite{Osipenko}, are required to provide
fully conclusive support for the Gauss model. However, we still may view 
the picture in Fig.~\ref{Fig3-test-Gauss-HERMES} as an encouraging indication 
in favor of the applicability of the Gauss model in SIDIS at HERMES.

%=== SECTION 2C: PARAMETERS FROM HERMES ==============================
\subsection{Determining parameters from HERMES, and cross check with CLAS}
\label{Subsec-2C:HERMES-fixing}

\begin{figure}[t!]
        \vspace{-0.7cm}
        \centering
        \includegraphics[width=6.5cm]{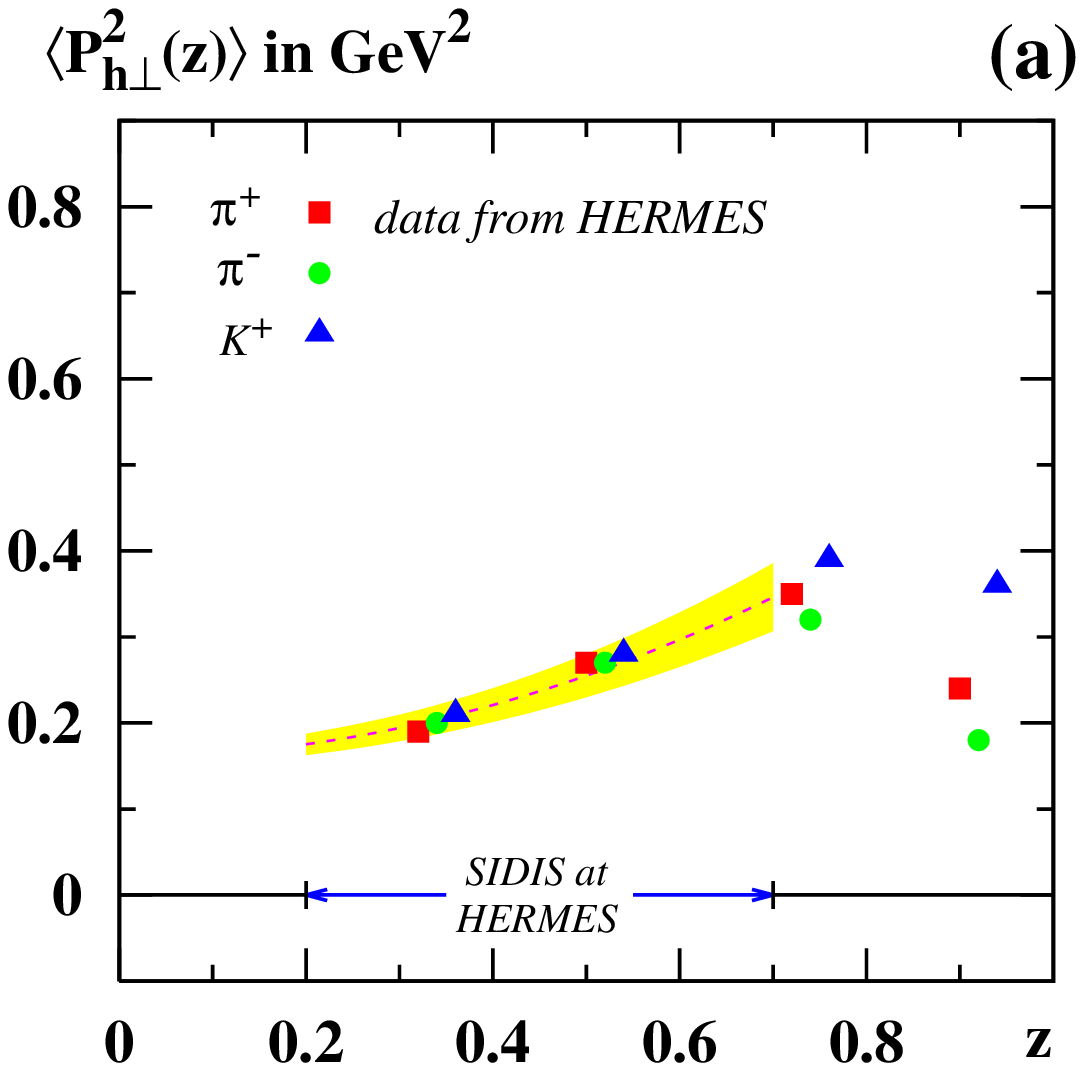}
        \includegraphics[width=6.5cm]{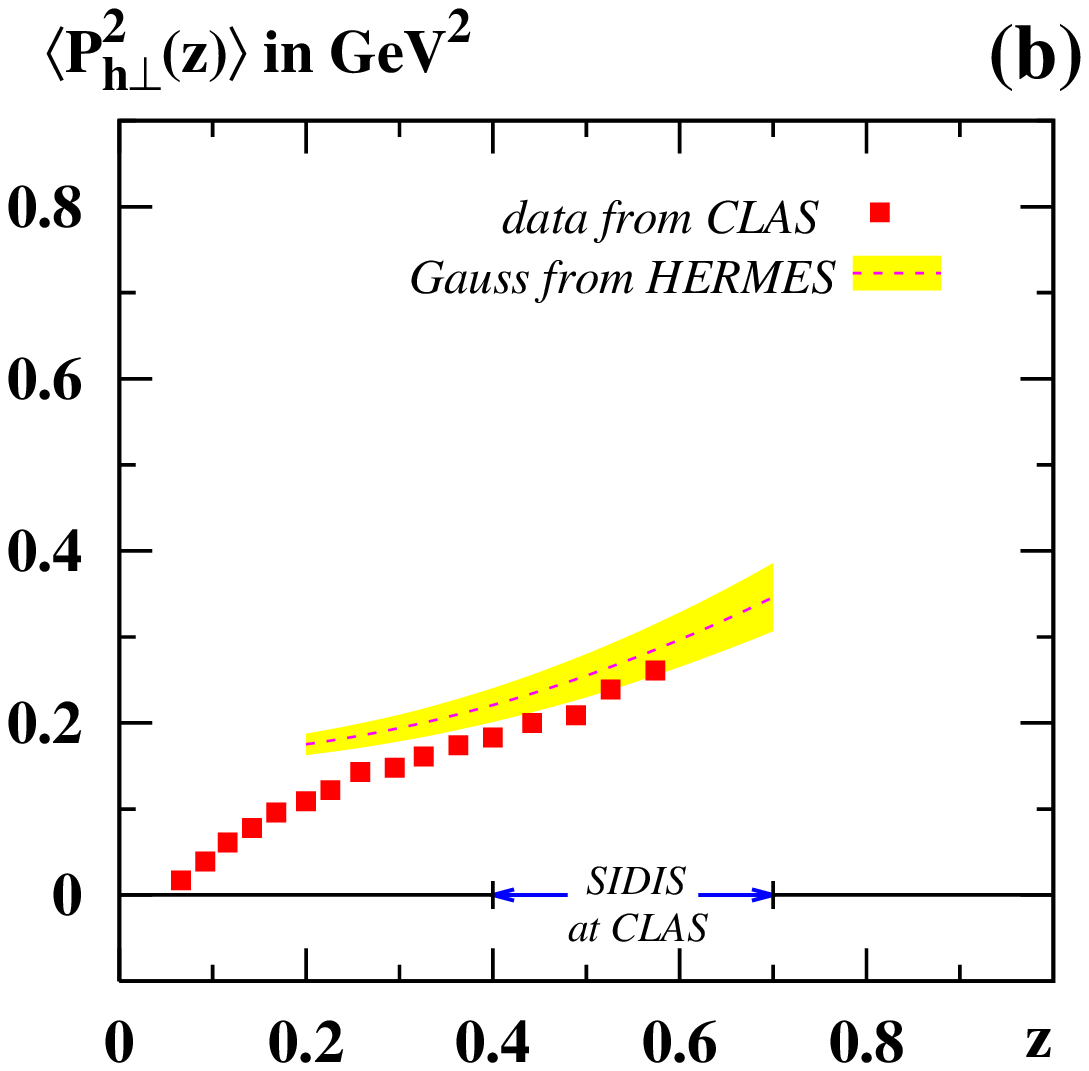}
        \vspace{-0.5cm}
        \caption{\label{Fig-04:HERMES-fixing-CLAS-cross-check}
        (a) $\la P^2_{h\perp}(z)\ra$ of hadrons in SIDIS off deuterium 
	at HERMES vs.~$z$ from \cite{Airapetian:2009jy}.
        The dotted line and the shaded region are the best fit and 
        its 1--$\sigma$ region from Eq.~(\ref{Eq:new-fit-HERMES}),
        see text.
        (b) $\la P^2_{h\perp}(z)\ra$ of $\pi^+$ in SIDIS off proton
	at CLAS vs.~$z$ \cite{Osipenko}. Dotted line (shaded region) 
        are the best fit (its 1--$\sigma$ region) to HERMES data from
        Fig.~\ref{Fig-04:HERMES-fixing-CLAS-cross-check}a, see text.}
        \end{figure}

The HERMES data \cite{Airapetian:2009jy} allow two further
insights. The first is that the $\la P_{h\perp}^2(z)\ra$ 
for $\pi^+$, $\pi^-$, $K^+$ are very similar in the SIDIS region,
i.e.,\ below the cut $z=0.7$ which excludes exclusive effects at HERMES, 
see Fig.~\ref{Fig-04:HERMES-fixing-CLAS-cross-check}a. Thus, there is no 
evidence of a strong flavor-dependence of the Gauss widths $\la p_T^2\ra$ 
and $\la K_T^2\ra$ at HERMES. The second insight is that the data in 
Fig.~\ref{Fig-04:HERMES-fixing-CLAS-cross-check}a allow to fix the Gauss 
widths  $\la p_T^2\ra$ and $\la K_T^2\ra$. A fit in the region $0.2 < z < 0.7$ 
yields 
\ba
      \la p_T^2\ra &=& (0.38\pm 0.06)\,{\rm GeV}^2\,, \nonumber\\
      \la K^2_T\ra &=& (0.16\pm 0.01)\,{\rm GeV}^2\,, 
      \label{Eq:new-fit-HERMES}\;\;\ea
with a $\chi^2$ per degree of freedom of 0.44. The best fit and its 1--$\sigma$ 
region are shown in Fig.~\ref{Fig-04:HERMES-fixing-CLAS-cross-check}a as
(respectively) dotted line and shaded region.

The values of the new fit in Eq.~(\ref{Eq:new-fit-HERMES}) are in good 
agreement with the results from \cite{Collins:2005ie,Anselmino:2005nn}
quoted in Eq.~(\ref{Eq:old-fit-pT2-KT2}), especially if one recalls
that those numbers have unestimated systematic uncertainties. This is 
the main point, in which our new result (\ref{Eq:new-fit-HERMES}) 
constitutes an improvement over the old numbers (\ref{Eq:old-fit-pT2-KT2}):
it has a well-estimated uncertainty.

In any case, the good news is that the numbers for the Gauss
widths from the works \cite{Collins:2005ie,Anselmino:2005nn} are
confirmed numerically. This means that, in this respect, phenomenological 
results for fixed target experiments obtained on the basis of those numbers
remain valid.

%------ BEGIN FIGURE 5: Phperp(z) at HERMES & CLAS -------------------
%------ END FIGURE 5 -------------------------------------------------

Finally, let us turn back to the CLAS data \cite{Osipenko} on the
$z$-dependence of $\la P^2_{h\perp}(z)\ra$ which refer to 
$\la Q^2\ra = 2.37\,{\rm GeV}^2$ which is similar to HERMES,
but $\la x\ra = 0.27$ which is substantially higher than HERMES.
In Sec.~\ref{Subsec-2A:Lesson-from-JLab} we refrained from using
these data to determine quantitatively Gauss model parameters
for reasons discussed there in detail. At this point it is
instructive to compare the fit result obtained from HERMES to
the CLAS data, see Fig.~\ref{Fig-04:HERMES-fixing-CLAS-cross-check}b.
Clearly, in the region above $z>0.4$ where the produced pions
are predominantly from current fragmentation, we observe a good
agreement. This indicates that it is the same  non-perturbative 
mechanism which generates intrinsic transverse momenta in the
two experiments.

%=== SECTION 2D: EMC AND CAHN ========================================
\subsection{Cahn effect at EMC}
\label{Subsec-2D:Cahn-effect+EMC-data}

In SIDIS of unpolarized leptons off unpolarized nucleons 
the cross section differential in the azimuthal angle of the produced 
hadrons, see Fig.~\ref{Fig-01:kinematics}, is --- for purely electromagnetic 
interactions --- given by \cite{Hagiwara:1982cq}
\be\label{Eq:unpol-SIDIS-general}
     \frac{\di^5\sigma(x,y,z,P_{h\perp},\phi)}{
       \di x\,\di y\,\di z\,\di P_{h\perp}^2\di\phi}=
     \frac{\di^4\sigma_{UU}(x,y,z,P_{h\perp})}{(2\pi)\,
       \di x\,\di y\,\di z\,\di P_{h\perp}^2\di\phi}
     \Biggl[1+\frac{(2-y)\sqrt{1-y}}{1+(1-y)^2}\,\cos(\phi)\;A_{UU}^{\cos\phi} 
             +\frac{1-y}{1+(1-y)^2}\,\cos(2\phi)\;A_{UU}^{\cos2\phi} \biggr]\,,
\ee
where $d^4\sigma_{UU}(x,y,z,P_{h\perp})$ denotes the differential cross section 
$d^5\sigma_{UU}(x,y,z,P_{h\perp},\phi)$ after averaging over $\phi$, as defined 
in Eq.~(\ref{Eq:sigma0}). $A_{UU}^{\cos\phi}$ and $A_{UU}^{\cos2\phi}$
in (\ref{Eq:unpol-SIDIS-general}) depend on $x$, $z$, $P_{h\perp}$
and are ratios of adequately defined structure functions.
The meaning of the subscript ``$UU$'' is as in (\ref{Eq:sigma0}), 
the superscript recalls the type of $\phi$-modulation.

At low $P_{h\perp}$, the observable $A_{UU}^{\cos\phi}$ is suppressed 
by $1/Q$, and in this sense a ``twist-3'' effect. 
At present it is not clear whether there is factorization in SIDIS 
at subleading twist \cite{Gamberg:2006ru,Bacchetta:2008xw}.
This $\cos\phi$-modulation is sometimes 
referred to as the Cahn effect. In Ref.~\cite{Cahn:1978se}
it was shown that the existence of intrinsic transverse parton momenta 
in the unpolarized distribution and fragmentation functions can generate
such a modulation. Later it became clear that, if one assumes factorization
and restricts oneself to a lowest order QCD (``tree-level'') description,
there are four different contributions to this asymmetry from several TMDs 
and $K_T$-dependent fragmentation functions 
\cite{Kotzinian:1994dv,Mulders:1995dh,Boer:1997nt,Bacchetta:2006tn}.
More precisely, in the notation of \cite{Bacchetta:2006tn} the asymmetry is given by
\ba
A_{UU}^{\cos\phi}&=&\frac{F_{UU}^{\cos\phi}}{F_{UU,T}}\;,\;\;\;\;
F_{UU,T} = {\cal C}\biggl[\;f_1 D_1 \;\biggr ]\,,\nonumber\\     
F_{UU}^{\cos\phi}  
	&=& \frac{2M}{Q}\,{\cal C}\biggl[
   	\frac{\vec{K}_T\cdot\vec{P}_{h\perp}}{m_hP_{h\perp}} 
	\biggl( x h\,H_{1}^{\perp } 
   	+ \frac{m_h}{M}\,f_1 \frac{\tilde{D}^{\perp }}{z}\biggr)
   	- \frac{\vec{p}_T\cdot\vec{P}_{h\perp}}{MP_{h\perp}} 
     	\biggl( x  f^{\perp } D_1
   	+ \frac{m_h}{M}\,h_{1}^{\perp } \frac{\tilde{H}}{z}\biggr)\biggr]\,,
        \label{Eq:FUUcosphi}\ea
where $m_h$ denotes the hadron mass, and the meaning of the convolution 
symbol is, for generic functions $f$ and $D$, 
\be\label{Eq:convol-integral}
	{\cal C}\biggl[\;\dots\;f\;D\biggr]=\int\di^2p_T^{ }\int\di^2K_T
	\;\delta^{(2)}(z\vec{p}_T^{ }+\vec{K}_T^{ }-\vec{P}_{h\perp}^{ })\;
        \dots\;\sum_a e_a^2\;x\,f^a(x,p_T)\;D^a(z,K_T)\;.
\ee

The original ``Cahn-effect-only'' description of $A_{UU}^{\cos\phi}$
\cite{Cahn:1978se} can be ``rederived'' from the TMD formalism
\cite{Mulders:1995dh,Boer:1997nt,Bacchetta:2006tn} under two conditions.
The first is that quark-gluon-quark correlators, which can be separated 
off by exploring QCD equations of motion, can be neglected
with respect to quark-quark correlators. The quality of such approximations was 
discussed in \cite{Avakian:2007mv,Metz:2008ib,Teckentrup:2009tk,Accardi:2009au}.
One of these approximations is $xf^{\perp q}\approx f_1^q$, 
and is supported by results from the bag model \cite{Avakian:2010br}.
The second condition is that in the experiment the type of hadron $h$
is not detected, i.e.,\  the observed hadron with momentum specified by 
$z$, $\phi$, $P_{h\perp}$ is a sum over pions, kaons, etc., and in such 
sums the Collins effect tends to cancel. Strictly speaking, 
an exact cancellation occurs only if one weighs the Collins function 
$H_1^\perp$ adequately, integrates over $0\le z\le1$, and sums 
over {\sl all} hadrons \cite{Schafer:1999kn}. 
However, string fragmentation models \cite{Artru:1995bh} and phenomenology
\cite{Vogelsang:2005cs,Efremov:2006qm,Anselmino:2007fs,Anselmino:2008jk}
indicate that cancellations occur in practice already if one sums over charged 
hadrons only, which are mostly $\pi^\pm$ in experiments such as EMC, where 
the asymmetry $A_{UU}^{\cos\phi}$ in charged hadron production was measured 
\cite{Aubert:1983cz,Arneodo:1986cf}.

%------ BEGIN FIGURE 6: AUUcosphi EMC --------------------------------
        \begin{wrapfigure}[28]{RD!}{9cm}
        \centering
        \includegraphics[width=8.5cm]{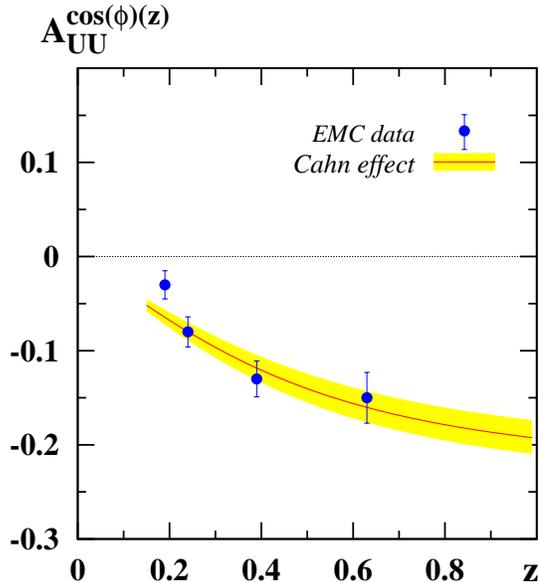}
        \caption{\label{Fig5-EMC}
        Azimuthal asymmetry $A_{UU}^{\cos\phi}$ in charged hadron 
        production vs.\ $z$. The data are from the
        EMC experiment \cite{Aubert:1983cz}. The theoretical
        curve is the ``Cahn-effect-only'' approximation for
        this observable, which is justified under certain 
        assumptions (see text), using the Gauss model 
        with parameters fixed from HERMES, Eq.~(\ref{Eq:new-fit-HERMES}).}
        \end{wrapfigure}
%------ END FIGURE 6 -------------------------------------------------

In fact, the EMC data on  $A_{UU}^{\cos\phi}$ 
\cite{Aubert:1983cz,Arneodo:1986cf} were used in 
\cite{Anselmino:2005nn} to determine in the 
``Cahn-effect-only'' approximation the Gauss model parameters quoted 
in Eq.~(\ref{Eq:old-fit-pT2-KT2}). It is therefore instructive to check,
whether the revised numbers from (\ref{Eq:new-fit-HERMES}) still give
a good description of these data.

In the ``Cahn-effect-only'' approximation, i.e.,\ neglecting in 
(\ref{Eq:FUUcosphi}) the ``pure twist-3'' tilde-functions and the 
Collins effect, we obtain for the asymmetry with the Gauss model 
(\ref{Eq:Gauss-f1},~\ref{Eq:Gauss-D1}) the result 
\be\label{Eq:AUUcosphiCahn}
    A_{UU, \rm Cahn}^{\cos\phi}(z)=
    -\;\frac{z\;\sqrt{\pi}\;\la p_T^2\ra}
            {\la Q\ra\sqrt{z^2\la p_T^2\ra + \la K_T^2\ra}} \;,    
\ee
where we assumed flavor-independent Gauss widths.
In the EMC \cite{Aubert:1983cz} experiment $280\,{\rm GeV}$ muons 
were scattered off protons and the covered kinematics was: $y<0.8$, 
$z>0.15$, $Q^2>10\,{\rm GeV}^2$ with about $\la Q\ra=4.8\,{\rm GeV}$,
and $160 < W^2/{\rm GeV}^2 < 360$. 
Also the cut $P_{h\perp}>200\,{\rm MeV}$ was imposed which is ignored in 
(\ref{Eq:AUUcosphiCahn}) for simplicity. Numerically it has a negligible
effect.
In this kinematics we obtain with the Gauss model parameters inferred
from HERMES, Eq.~(\ref{Eq:new-fit-HERMES}), the result shown in 
Fig.~\ref{Fig5-EMC}. We observe a very good agreement.

At this point, this not only demonstrates the compatibility of the EMC 
and HERMES data. Since we have fixed the details of the Gauss model in 
an independent experiment, the excellent agreement we observe in 
Fig.~\ref{Fig5-EMC} can also be read in opposite direction. 
The approximations we used in order to relate $A_{UU}^{\cos\phi}$
to the Cahn effect, namely the cancellation of the Collins effect 
in charged hadron production and the neglect of tilde-terms, are 
justified --- within the experimental error bars and the theoretical
uncertainty of our study. In this sense, the EMC data on 
$A_{UU}^{\cos\phi}$ support the ``Wandzura-Wilczek-type''
approximations discussed critically in 
\cite{Avakian:2007mv,Metz:2008ib,Teckentrup:2009tk,Accardi:2009au}.
(However, we shall come back to this point at the end of 
Sec.~\ref{Sec-4:s-dependence}.)

To draw an intermediate summary, in the Sections
\ref{Subsec-2A:Lesson-from-JLab}--\ref{Subsec-2D:Cahn-effect+EMC-data}
we have seen that data from EMC \cite{Aubert:1983cz,Arneodo:1986cf},
Jefferson Lab \cite{Osipenko,Mkrtchyan:2007sr} and HERMES 
\cite{Airapetian:2009jy} support the Gauss model with 
flavor- and $x$- or $z$-independent Gauss widths.
In the future one may need to refine the description by allowing for flavor- 
and $x$- or $z$-dependent Gauss widths, when more precise data will 
make it necessary to introduce and possible to constrain further parameters.

The Gauss model has an important principle limitation though.
It may work only if the transverse momenta of the produced hadrons
are of the order of magnitude of the hadronic scale, i.e.,\ much
smaller than the hard scale in the process. This condition is
fulfilled in the case of EMC, Jefferson Lab, and HERMES data
\cite{Aubert:1983cz,Arneodo:1986cf,Osipenko,Mkrtchyan:2007sr,Airapetian:2009jy}
where typically $\la P_{h\perp}\ra \simeq 0.4\,{\rm GeV} \ll \la Q\ra=$
2--5$\,{\rm GeV}$ in these experiments.
When transverse hadron momenta become substantially larger
than the hadronic scale or even become so large that they set
the hard scale in the process, one does not deal with 
non-perturbative intrinsic $p_T$ anymore, but can
apply perturbative QCD \cite{Georgi:1977tv}.
SIDIS data from high energy experiments, for example E665 at Fermilab 
\cite{Adams:1993hs} or ZEUS at DESY \cite{Breitweg:2000qh,Chekanov:2006gt},
are sensitive to such perturbative $p_T$-effects.
In practice, in order to describe these data it is necessary to
include both, perturbative and non-perturbative effects.
We refer to the  pioneering study \cite{Konig:1982uk},
see also the recent work \cite{Anselmino:2006rv}.

\newpage

%=== SECTION 2E: CAHN & BOER-MULDERS =================================
\subsection{Cahn and Boer-Mulders effect, and new data}
\label{Subsec-2E:Cahn-BM}

It would be interesting to repeat the analysis presented in the previous 
section with recent results on $A_{UU}^{\cos\phi}$ from Jefferson Lab
\cite{Osipenko,Mkrtchyan:2007sr} (final), HERMES and COMPASS 
\cite{Giordano:2009hi,Kafer:2008ud} (preliminary data).
Since $\la Q\ra$ in these experiments is smaller compared to EMC, 
one would expect this subleading-twist observable to be there larger 
than at EMC.

A careful comparison of the Cahn effect prediction for 
$A_{UU}^{\cos\phi}$ to these new data will be instructive
for two reasons. First, at Jefferson Lab, COMPASS, HERMES the 
hadrons are identified, i.e.,\ the Collins effect in 
(\ref{Eq:FUUcosphi}) does not cancel out --- in contrast to EMC, 
where we could argue it may (at least approximately) cancel out. 
Second, the new or forthcoming data are far more precise compared 
to EMC, such that deviations of data from the ``Cahn-effect-only''
approximation will have a chance to pop up more easily, and 
provide insights on quark-gluon correlations.
In view of the preliminary status of the HERMES and COMPASS data 
\cite{Giordano:2009hi,Kafer:2008ud}, however, we shall not pursue
this study here.

Let us also comment on the Boer-Mulders effect responsible for the 
$\cos(2\phi)$-modulation in (\ref{Eq:unpol-SIDIS-general}) and given by
\be
A_{UU}^{\cos2\phi} = \frac{F_{UU}^{\cos2\phi}}{F_{UU,T}}
                   + {\cal O}\biggl(\frac{M^2}{Q^2}\biggr)\;,\;\;\;\;
F_{UU}^{\cos2\phi} = {\cal C}\biggl[
   	  \frac{2(\vec{K}_T\cdot\vec{P}_{h\perp})
                 (\vec{p}_T\cdot\vec{P}_{h\perp})
                -(\vec{K}_T\cdot\vec{p}_T)\,P^2_{h\perp}}{Mm_hP^2_{h\perp}}\;
              h_1^\perp\,H_{1}^{\perp} \biggr] \,, \label{Eq:FUUcos2phi}\ee
with $F_{UU,T}$ and the convolution integral as defined 
in (\ref{Eq:FUUcosphi},~\ref{Eq:convol-integral}).
The Boer-Mulders function $h_1^\perp$ \cite{Boer:1997nt} is one of the 
so-called ``naively time-reversal odd'' (T-odd) TMDs 
\cite{Brodsky:2002cx,Collins:2002kn,Belitsky:2002sm}.
The observable was discussed in
\cite{Oganesian:1997jq,Gamberg:2003ey,
Barone:2005kt,Barone:2006ws,Gamberg:2007wm,Zhang:2008ez,Barone:2008tn,Barone:2009hw}.

In Eq.~(\ref{Eq:FUUcos2phi}) we have indicated the power corrections
modulo which the factorization theorem is formulated, e.g.\
\cite{Efremov:1980kz,Collins:1981uk,Collins:1984kg,Ji:2004wu}. 
Only if such corrections
are sufficiently small the factorization approach is justified, and can
develop its predictive power with universal non-perturbative objects 
\cite{Collins:2004nx}, though the concept of universality had to be
extented in order to accommodate T-odd TMDs
\cite{Brodsky:2002cx,Collins:2002kn,Belitsky:2002sm}, which we will
discuss in more detail below in Sec.~\ref{Sec-3C:Cahn-BM-DY}.

In general such non-factorizable corrections are theoretically not under 
control and must be excluded experimentally by studying the $Q$-behavior 
of the cross section or asymmetry. In the case of $A_{UU}^{\cos2\phi}$, 
however, the parton model provides a way to estimate one of the possible 
power corrections in  (\ref{Eq:FUUcos2phi}). Namely, the Cahn effect 
\cite{Cahn:1978se}  yields
\be\label{Eq:AUUcos2phiCahn}
    A_{UU, \rm Cahn}^{\cos2\phi}(z)=
    -\;\frac{\la p_T^2\ra}{\la Q^2\ra}\;
    \frac{z^2\;\la p_T^2\ra}{z^2\la p_T^2\ra+\la K_T^2\ra}\;,   
\ee
if we assume the Gauss model with flavor-independent Gauss widths.
We stress that this is only one among many {\sl non-factorizable} 
power-corrections to the $\cos(2\phi)$-modulation.
Ideally, one should choose the cuts in $Q$ in the experiment such
that one does not feel the Cahn- (\ref{Eq:AUUcos2phiCahn}) or
any other $1/Q^2$-power correction in (\ref{Eq:FUUcos2phi}).

In practice, this is difficult. At HERMES and COMPASS the DIS cut 
$Q^2>1\,{\rm GeV}^2$ is imposed which leads to 
$\la Q^2\ra\simeq2.5\,{\rm GeV}^2$. 
At Jefferson Lab similar $\la Q^2\ra$ (but lower $W$) is achieved, 
in spite of lower beam energies, because there one can afford to choose 
the cut $Q^2>2\,{\rm GeV}^2$ \cite{Mkrtchyan:2007sr} thanks to the high 
luminosity. In such kinematics, at let us say $z=0.5$, the Cahn-power 
correction (\ref{Eq:AUUcos2phiCahn}) is, with Gauss widths from 
(\ref{Eq:new-fit-HERMES}), about
\be\label{Eq:AUUcos2phiCahn-estimate}
    A_{UU, \rm Cahn}^{\cos2\phi} \simeq -{\cal O}(5\,\%)
    \;.\ee
Let us compare this number to what one may expect from the leading
twist Boer-Mulders contribution in (\ref{Eq:FUUcos2phi}). If we 
generously allow the Boer-Mulders function to saturate its positivity bound 
$\frac{p_T}{M}|h_1^{\perp a}(x,p_T)|\le f_1^a(x,p_T)$ \cite{Bacchetta:1999kz},
then 
\be
       |F_{UU}^{\cos2\phi}(x,z)| \le 
       {\cal C}\biggl[\omega_{\rm BM}\;
               f_1\;\frac{K_T}{m_h}\,|H_1^{\perp}| \biggr]
               = \alpha_{\rm G}
               \sum\limits_ae_a^2 \,x\,f_1^a(x)\,|H_1^{\perp(1/2)a}(z)| \,,
\ee
where $\omega_{\rm BM}=2(\hat{K}_T\cdot\hat{P}_{h\perp})
(\hat{p}_T\cdot\hat{P}_{h\perp})-(\hat{K}_T\cdot\hat{p}_T)$, with the ``hat'' 
denoting unit vectors and the transverse (1/2)-moment defined as
$H_1^{\perp(1/2)a}(z)=\int\di^2K_T\frac{K_T}{m_h}H_1^{\perp a}(z,K_T)$.
The numerical factor $\alpha_{\rm G}={\cal O}(1)$ contains the
dependence on the model for the transverse parton momenta.
Phenomenological studies indicate that 
$|H_1^{\perp(1/2)a} / D_1^a| = {\cal O}(10\%)$ around $z=0.5$. Thus, the 
leading-twist Boer-Mulders contribution in (\ref{Eq:FUUcos2phi})
can be expected to be at most
\be
       |A_{UU,\rm BM}^{\cos2\phi}| \lesssim {\cal O}(10\%)\:.
\ee
Hence, the power correction estimated on the basis of the Cahn effect
in (\ref{Eq:AUUcos2phiCahn-estimate}) is substantial in this kinematics.
Studies of the Boer-Mulders effect in SIDIS 
\cite{Oganesian:1997jq,Gamberg:2003ey,
Barone:2005kt,Barone:2006ws,Gamberg:2007wm,Zhang:2008ez,Barone:2008tn}
and first analyses of preliminary data \cite{Barone:2009hw} yield 
similar results.
We conclude therefore that, unless it will be possible to suppress 
such power corrections by applying sufficiently large cuts in $Q$,
or separate experimentally the Boer-Mulders part and power corrections 
by their different $Q$-behavior, the task of gaining insights on the 
Boer-Mulders function from present SIDIS data will be highly demanding 
--- and feasible only on a basis of a thorough quantitative understanding 
of intrinsic transverse momentum effects in unpolarized distribution 
and fragmentation functions, which is what this work aims at. 
Interestingly, as we shall see in Section~\ref{Sec-3:DY}, 
the situation is less demanding for the available DY data. 

\newpage
%=== SECTION 6: DRELL-YAN ============================================
\section{Intrinsic transverse momenta in Drell-Yan}
\label{Sec-3:DY}

In this Section we discuss the DY process \cite{Drell:1970wh}, i.e.,\ 
the inclusive production of large-invariant-mass $\mu^+\mu^-$ pairs 
in hadron-hadron collisions $h_1h_2\to\mu^+\mu^-X$, first observed 
in 1970 at the Brookhaven AGS \cite{Christenson:1970um}, for reviews see 
\cite{Stirling:1993gc,McGaughey:1999mq,Reimer:2007iy,
Tangerman:1994eh,Arnold:2008kf}.
Let $p_{1,2}$ and $k_{1,2}$ denote (respectively) the momenta of the incoming 
hadrons $h_{1,2}$ and the outgoing lepton pair. The kinematics of the process 
is described by the center of mass energy square $s$, 
invariant mass of the lepton pair $Q$, rapidity $y$ or the Feynman variable 
$x_F$, and the variable $\tau$ which are defined as
\ba
&&	s    =      (p_1+p_2)^2 	\,,\;\;\; 
	Q^2  =      (k_1+k_2)^2 	\,,\;\;\; 
   	y    =      \frac12\,\ln\frac{p_2\cdot(k_1+k_2)}{p_1\cdot(k_1+k_2)} 
	     \equiv \frac12\,\,\ln\frac{x_1}{x_2}\,,\;\;\;\nonumber\\
&&	x_F  \equiv x_1 - x_2	\,,\;\;\;
	\tau \equiv \frac{Q^2}{s} = x_1 x_2	\,.
\ea
The three-momentum of the virtual photon $\vec{q}=\vec{k}_1+\vec{k}_2$ 
can be decomposed in the center of mass frame of the incoming hadrons into 
a longitudinal and transverse component with respect to the collision axis as 
$\vec{q}=(q_L,\vec{q}_T)$. In this frame $y=\frac12\,{\rm ln}(Q+p_L)/(Q-p_L)$ 
and $x_F = 2q_L/\sqrt{s}$. 

In the parton model $x_i$ denotes the fraction of the hadron momentum $p_i$ 
carried by (respectively) the annihilating parton or anti-parton, and the 
cross section differential in $x_1$, $x_2$ and $q_T=|\vec{q}_T|$ is given by
(we write $f_1^a(x_i)\equiv {f_1^a}{ }^{/h_i}(x_i)$ for brevity, i.e.,\  the 
parton distributions in the possibly distinct hadrons $h_i$ are unambiguously
labelled by the $x_i$)
\be\label{Eq:DY-qT-00}
	\frac{\di^3\sigma_{UU}}{\di x_1\,\di x_2\,q_T\,\di q_T} 
	= \frac{4\pi\alpha^2}{9Q^2}\,\sum_{a=u,d,\bar u,\dots}e_a^2
	\int\limits_0^{2\pi}\di\phi\int\di^2 p_{1T} \int\di^2p_{2T}
	\,\delta^{(2)}(\vec{p}_{1T}+\vec{p}_{2T}-\vec{q}_T)
	f^a_1(x_1,p_{1T})f^{\bar a}_1(x_2,p_{2T})\;.
\ee
For an accurate, quantitative description of the $q_T$-dependence of the 
DY cross section it is necessary to go beyond the parton model expression 
(\ref{Eq:DY-qT-00}) and use the Collins-Soper-Sterman formalism 
\cite{Collins:1984kg}.

However, it turns out that a very useful effective description
can be obtained by assuming the Gauss model (\ref{Eq:Gauss-f1}).
Inserting (\ref{Eq:Gauss-f1}) into (\ref{Eq:DY-qT-00}), 
and assuming that the Gauss widths ($\la p_{1T}^2\ra $ in hadron 1, 
and $\la p_{2T}^2\ra $ in hadron 2) are flavor-independent, yields
\be\label{Eq:DY-qT-01}
	\frac{\di^3\sigma_{UU}}{\di x_1\,\di x_2\,\di q_T} 
	= \frac{4\pi\alpha^2}{9Q^2}\,\sum_{a=u,d,\bar u,\dots}e_a^2
	f^a_1(x_1)f^{\bar a}_1(x_2)\;
	2q_T\,\frac{\exp(-q_T^2/\kappa_{DY}^2)}{\kappa_{DY}^2}\;,\;\;\;
        \kappa_{DY}^2 \equiv \la p_{1T}^2\ra  + \la p_{2T}^2\ra  \;.
\ee
We define the mean transverse momentum $\la q_T\ra$ of the muon pair,
and the mean transverse momentum square $\la q_T^2\ra$ as 
\ba\label{Eq:DY-qT-01a}
   \la q_T^2\ra  =  
   \biggl\la \frac{q_T^2\,\di\sigma_{UU}/\di q_T}{\di\sigma_{UU}/\di q_T}\biggr\ra
   &\stackrel{\rm Gauss}{=}&  \kappa_{DY}^2 \,,\nonumber\\
   \la q_T\ra  =  
   \biggl\la \frac{q_T\,\di\sigma_{UU}/\di q_T}{\di\sigma_{UU}/\di q_T}\biggr\ra
   &\stackrel{\rm Gauss}{=}& \frac{\sqrt{\pi}}{2}\,\sqrt{\kappa_{DY}^2}\;.
\ea
When integrating (\ref{Eq:DY-qT-01}) over $q_T$ the transverse momentum 
model dependence drops out and one obtains
\be\label{Eq:DY-qT-02}
	\frac{\di^2\sigma_{UU}}{\di x_1\,\di x_2} 
	= \frac{4\pi\alpha^2}{9Q^2}\,\sum_{a=u,d,\bar u,\dots}e_a^2
	f^a_1(x_1)f^{\bar a}_1(x_2)\;.
\ee
Using for $f_1^a(x)$ leading order parametrizations from DIS, the 
leading order QCD formula (\ref{Eq:DY-qT-02}) under\-estimates data by 
a (``K''-)factor $\sim(1.1$-$1.7)$ depending on $Q^2$ (and moderately 
also on $x_{1,2}$) \cite{Stirling:1993gc}.
 
One gets the overall normalization of the DY cross section correct
only by including higher order QCD corrections.
Throughout we shall work here in leading order. It is a remarkable observation 
that, whenever corrections to leading order DY cross sections were considered,
absolute cross sections were found to be considerably altered (typically enhanced), 
but ratios of observables in DY were found to be only moderately altered. 
This is in particular the case for collinear double spin asymmetries, 
for example the double transverse spin asymmetry $A_{TT}\propto$
$\sum_ae_a^2h_1^a(x_1)h_1^{\bar a}(x_2)/\sum_ae_a^2f_1^a(x_1)f_1^{\bar a}(x_2)$
\cite{Vogelsang:1992jn,Contogouris:ws,Ratcliffe:2004we,Shimizu:2005fp,Barone:2005cr,Kawamura:2007ze}.

\newpage
%=== SECTION 3A: TESTING GAUS IN DY ==================================
\subsection{Testing the Gauss model in DY}
\label{Subsec-3A:tesing-Gauss-in-DY}

%------ BEGIN FIGURE 7: <qT>^2 vs <qT^2> IN DY ------------------------
 \begin{figure}[b!]
 \centering
%\begin{wrapfigure}[27]{RD}{10cm}

	\includegraphics[width=13cm]{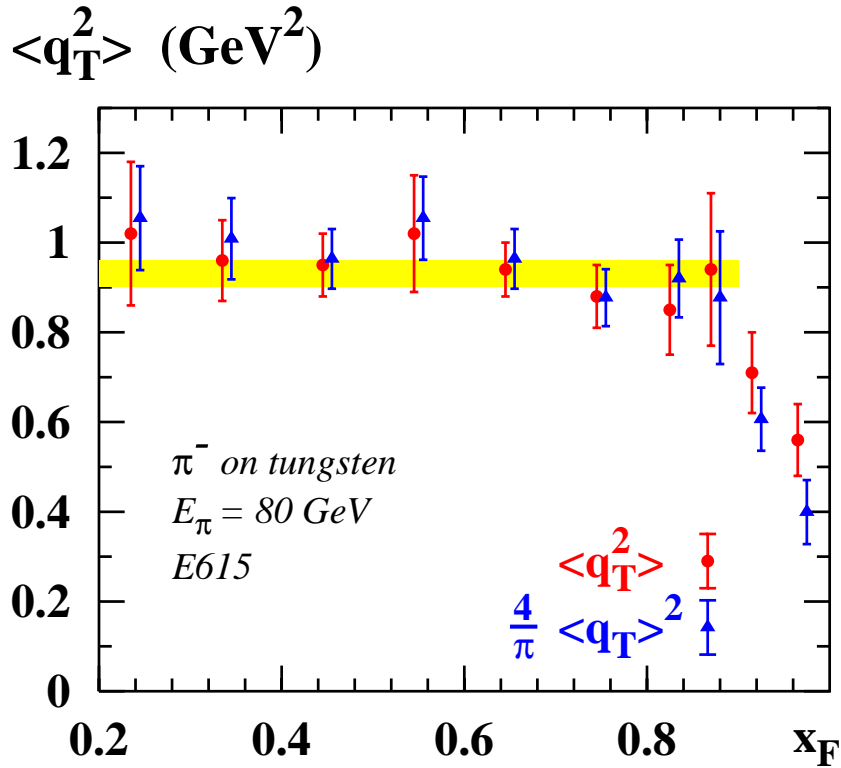}
        \caption{\label{FigUV:DY-qT-qT2}
	The mean dimuon transverse momentum square $\la q_T^2\ra$ 
  	vs.\ $x_F$ as measured in the Fermilab E615 experiment 
	\cite{Palestini:1985zc}.
	The data points for $\la q_T^2\ra$ are marked by circles,
	the data points for $\frac{4}{\pi}\la q_T\ra^2$ are marked by triangles.
	Both quantities are predicted to be equal in the Gauss model, 
	see Eq.~(\ref{Eq:qT-qT2-in-Gauss}),
	which is the case within the statistical accuracy of the data.}
%\end{wrapfigure}
 \end{figure}
%------ END FIGURE 7 -------------------------------------------------

The Gauss model unambiguously connects the average lepton pair transverse 
momenta and their squares. In fact, from (\ref{Eq:DY-qT-01a}) we obtain
\be\label{Eq:qT-qT2-in-Gauss}
	\la q_T^2\ra \stackrel{\rm Gauss}{=} \frac{4}{\pi}\;\la q_T\ra^2 \;,
\ee
which is a parameter-free prediction of the Gauss model, 
c.f.\ Eq.~(\ref{Eq:Gauss-predict}) in SIDIS.
Data allowing to check (\ref{Eq:qT-qT2-in-Gauss}) were reported 
in \cite{Palestini:1985zc}, and are shown in Fig.~\ref{FigUV:DY-qT-qT2}.
Several interesting conclusions can be drawn from Fig.~\ref{FigUV:DY-qT-qT2}.
First, it demonstrates that in the range $0.2 < x_F < 1$ covered in 
the experiment \cite{Palestini:1985zc} the relation (\ref{Eq:qT-qT2-in-Gauss})
is satisfied within the accuracy of the data.
Second, it demonstrates that for $0.2 < x_F < 0.9$ both $\la q_T^2\ra$ 
and $\la q_T\ra$ are $x_F$-independent within error bars. Therefore,
we conclude that for $x_F<0.9$ the Gauss model with $x$-independent 
Gauss widths provides an excellent description for the $q_T$-dependence 
of the DY process, which is useful at least for the description of data 
with an accuracy comparable to that of \cite{Palestini:1985zc}.
More precise data may reveal deviations from the Gauss Ansatz. 

It should be noticed that the parton model is not adequate 
at large $x_F$, where higher twist effects are expected to be relevant
\cite{Berger:1979du}.
Interestingly, the Gauss model relation (\ref{Eq:qT-qT2-in-Gauss})
holds even in the region $x_F>0.9$ within error bars,
see Fig.~\ref{FigUV:DY-qT-qT2}.

In Ref.~\cite{Badier:1982zb} $\la q_T^2\ra$ was measured for DY lepton pairs
induced by various hadrons on a platinum target. Within experimental accuracy 
it was found that $\la q_T^2\ra_{\pi^-Pt}\approx\la q_T^2\ra_{\pi^+Pt}$, etc.
Since these reactions are dominated by differently flavored valence quarks,
one may conclude that it is indeed approximately justified to use
flavor-independent Gauss widths. It is also important to notice, 
that the dependence on the nuclear target \cite{Bordalo:1987cs}
is moderate for transverse dilepton momenta $q_T\lesssim 3\,{\rm GeV}$,
i.e.,\ nuclear effects are not relevant.

The comparison in Fig.~\ref{FigUV:DY-qT-qT2} is very instructive, 
and allowed us to draw many interesting conclusions.
Yet, it is not a proof that the Gauss model really works in DY.
For that it is necessary to consider data on cross sections,
which we will do in the next section.

\newpage
%=== SECTION 3B: qT DEPENDENCE OF DY CROSS SECTION ===================
\subsection{$q_T$-dependence of the cross section} 
\label{Subsec-3B:sigma-qT-DY}

%------ BEGIN FIGURE 8: DY cross sections vs. qT ---------------------- 
\begin{figure}[b!]
\begin{center}
\includegraphics[height=7cm]{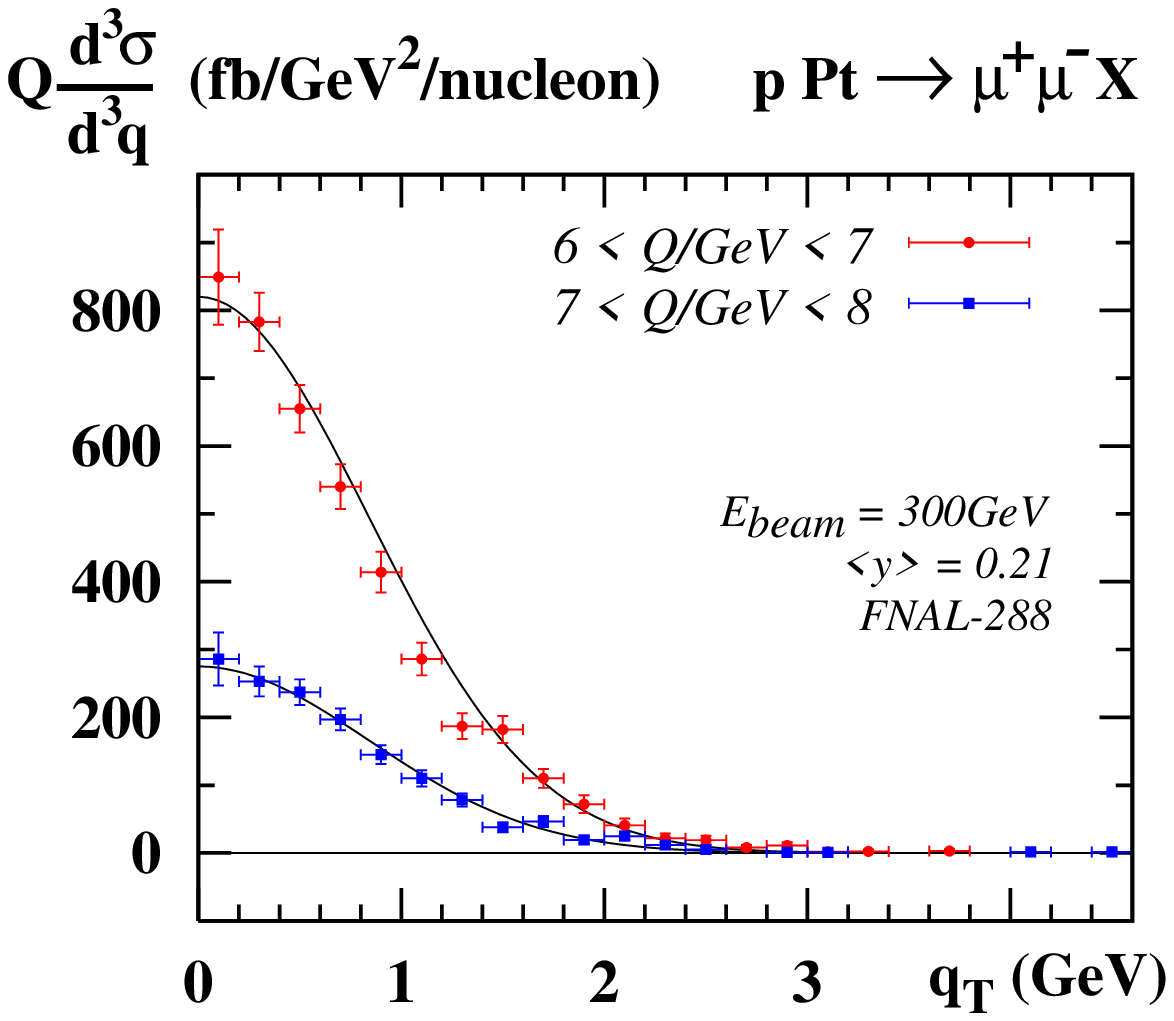}
\includegraphics[height=7cm]{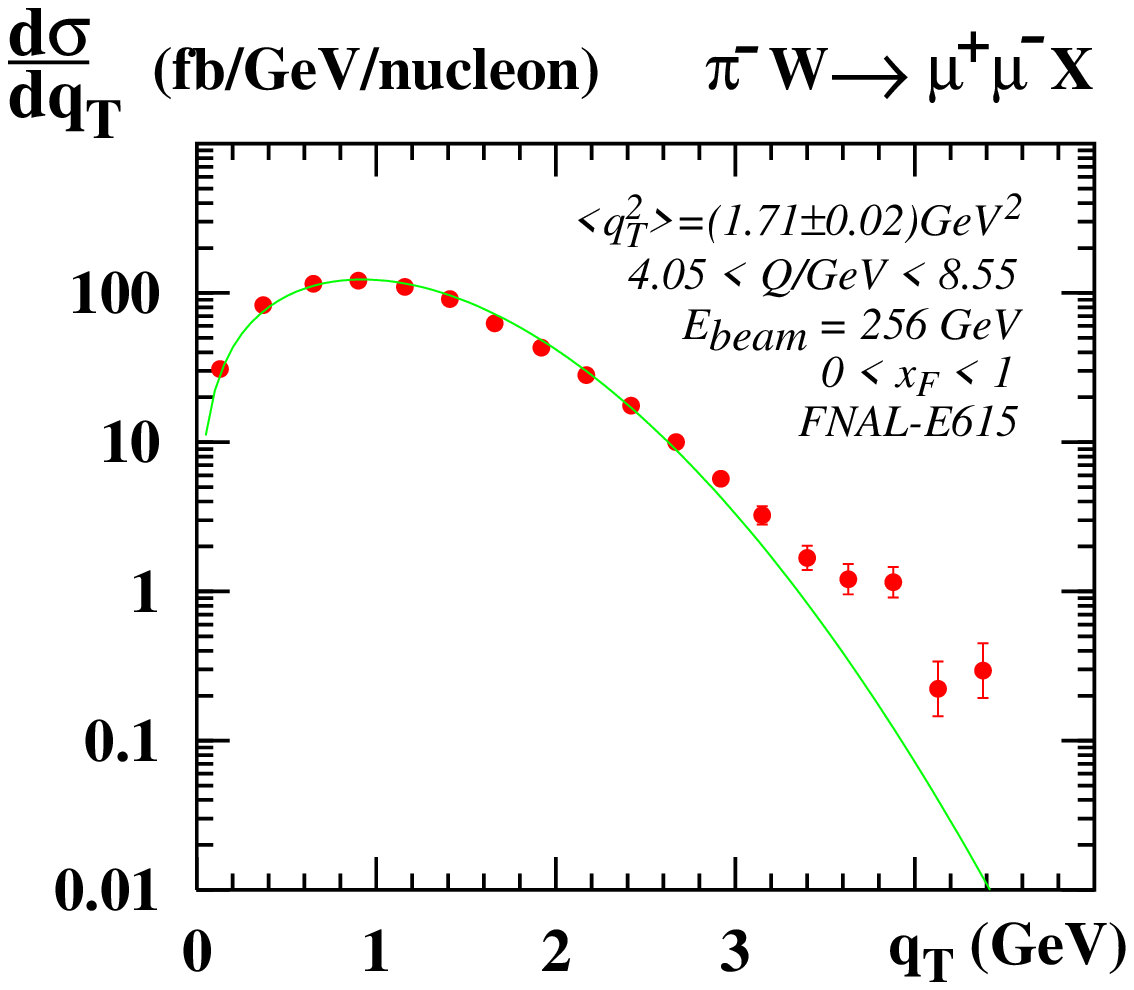}
\end{center}
        \vspace{-0.3cm}
        \caption{\label{FigXY:DY-qT}
	Left: 
        The invariant differential cross section $Q\frac{\di^3\sigma}{\di^3q}$ 
	for $p\;Pt\,\to\mu^+\mu^- X$ at $\la y\ra = 0.21$ for two different 
	$Q$-bins from FNAL-288 \cite{Ito:1981XX}.
	The Gauss model, Eq.~(\ref{Eq:DY-qT-03}), provides a good description
	of the data for $\la q_T^2\ra = 1.4\,{\rm GeV}^2$.\newline
	Right: 
        The differential cross section $\frac{\di\sigma}{\di q_T}$  
	for $\pi^-W\,\to\mu^+\mu^- X$ from FNAL-E615 \cite{Conway:1989fs}.
	Here the Gauss model, Eq.~(\ref{Eq:DY-qT-03}), provides a good 
	description of the data up to $q_T \lesssim 3\,{\rm GeV}$ with 
	$\la q_T^2\ra = 1.7\,{\rm GeV}^2$.}
\end{figure}
%-------- END FIGURE 8. ----------------------------------------------- 

The $q_T$-dependence of the cross sections allows to fix the parameters of 
the Gauss model.
It is convenient to rewrite the cross section  (\ref{Eq:DY-qT-01}) as
(here $\sigma_{UU}$ denotes the total $q_T$-integrated unpolarized DY 
cross section)
\be\label{Eq:DY-qT-03}
	\frac{\di\sigma_{UU}(q_T)}{\di q_T} =\sigma_{UU}\; 
	\frac{2\,q_T\exp(-q_T^2/\kappa_{DY}^2)}{\kappa_{DY}^2}
	\,,\;\;\;\mbox{or}\;\;\;\;
	Q\frac{\di\sigma_{UU}(q_T)}{\di^3 q} = 
	\frac{2}{\sqrt{s}}\frac{\di\sigma_{UU}}{\di x_F}\;
	\frac{\exp(-q_T^2/\kappa_{DY}^2)}{\kappa_{DY}^2} \;.
\ee

Fig.~\ref{FigXY:DY-qT} (left) shows FNAL-288 data on the $q_T$-dependence of 
the invariant differential cross section  $Q\frac{\di^3\sigma}{\di^3q}$ 
for $\mu^+\mu^-$ production from scattering a $300\,{\rm GeV}$ proton beam off 
platinum at $\la y\ra = 0.21$ for two different  bins in $Q$ \cite{Ito:1981XX}.
Clearly, the Gauss model provides a good description of the data for
\be\label{Eq:DY-qT2-in-p-p}
     \kappa_{DY}^2 \stackrel{\rm here}{=} 
     2\;\la p_{\mbox{\tiny $N$}T}^2\ra  = 1.4\,{\rm GeV}^2\,,
\ee
where $\la p_{\mbox{\tiny $N$}T}^2\ra$ denotes the mean parton momentum 
square in the nucleon.

The limitations of the Gauss Ansatz become more evident by 
using a logarithmic scale for the cross section. We do this in 
Fig.~\ref{FigXY:DY-qT}b that shows FNAL-E615 data \cite{Conway:1989fs}
for the differential cross section  $\frac{\di\sigma}{\di q_T}$ for 
$\mu^+\mu^-$ production from scattering a $256\,{\rm GeV}$ $\pi^-$ beam on a 
tungsten target for $0<x_F<1$ and $4.05\,{\rm GeV} < Q < 8.55\,{\rm GeV}$.
The Gauss model is applicable for $q_T \lesssim 3\,{\rm GeV}$ with 
the parameter 
\be\label{Eq:DY-qT2-in-p-pi}
     \kappa_{DY}^2\stackrel{\rm here}{=} 
     \la p_{\pi T}^2\ra + \la p_{\mbox{\tiny $N$}T}^2\ra  = 1.7\,{\rm GeV}^2\,,
\ee
where $\la p_{\pi T}^2\ra$ denotes the mean parton momentum square in the 
pion. 

The result (\ref{Eq:DY-qT2-in-p-pi}) corresponds precisely to the experimental 
average $\la q_T^2\ra = (1.71\pm 0.02)\,{\rm GeV}^2$  \cite{Conway:1989fs}. 
This is so because the 
differential cross section in the region $0\le q_T \lesssim 3\,{\rm GeV}$
(where the Gauss Ansatz is applicable) is about two orders of magnitude
larger than in the region $q_T \gtrsim 3\,{\rm GeV}$ where it is not.
Therefore the mean transverse momentum square is dominated by that region of $q_T$ 
where the Gauss model works. This means that in practice,
if the Ansatz works, the respective Gauss widths can be read 
off directly from the experimental results for $\la q_T^2\ra$.

Since the kinematics is comparable in the FNAL-288 and FNAL-E615 experiments, 
we can use (\ref{Eq:DY-qT2-in-p-p},~\ref{Eq:DY-qT2-in-p-pi}) to deduce
\be\label{DY-fix-widths}
        \la p_{\mbox{\tiny $N$}T}^2\ra  = 0.7\,{\rm GeV}^2\, , \;\;\;
        \la p_{\pi T}^2\ra  = 1.0 \,{\rm GeV}^2
        \;\;\;{\rm at}\;\;\;\sqrt{s} \sim 23\,{\rm GeV}.
\ee
Thus, transverse parton momenta appear to be larger in the pion than in 
the nucleon, as already noticed in \cite{Falciano:1986wk}.
The nucleon width in (\ref{DY-fix-widths}) is in good agreement with the 
value $\la p_{\mbox{\tiny $N$}T}^2\ra\equiv 1/\beta_0^2 = 0.64\,{\rm GeV}^2$
obtained at $\sqrt{s}\simeq 20\,{\rm GeV}$ in \cite{D'Alesio:2004up}.

%=== SECTION 3C: CAHN IN DY ==========================================
\subsection{Cahn and Boer-Mulders effect in DY}
\label{Sec-3C:Cahn-BM-DY}

Next we turn our attention to the azimuthal dependence of the DY
cross section in unpolarized hadron collisions, which was measured
in various experiments 
\cite{Badier:1981ti,Palestini:1985zc,Falciano:1986wk,Guanziroli:1987rp,Conway:1989fs,Zhu:2006gx,Zhu:2008sj}.
The general expression for the angular differential cross section 
in the Collins-Soper frame \cite{Collins:1977iv} is traditionally 
written as
\be\label{Eq:sigma-azimuthal-DY}
     \frac{1}{\sigma}\,\frac{\di\sigma}{\di\Omega}
     = \frac{3}{4\pi}\,\frac{1}{\lambda+3}\biggl(1+\lambda\cos^2\theta
     + \mu\sin(2\theta)\,\cos\phi + \frac{\nu}{2}\,\sin^2\theta\,\cos(2\phi)\biggr)\:.
\ee
In collinear QCD factorization to ${\cal O}(\alpha_s)$ the coefficients 
$\lambda$, $\mu$, $\nu$ are expected to obey the Lam-Tung relation 
\cite{Lam:1978pu,Collins:1978yt}
\be\label{Eq:Lam-Tung}
     2\nu+\lambda = 1\:,
\ee
which is preserved as a good approximation at ${\cal O}(\alpha_s^2)$
\cite{Mirkes:1994dp}.
But in experiments with pion beams on nuclear targets
the relation (\ref{Eq:Lam-Tung}) was found to be strongly violated:
while $\lambda={\cal O}(1)$ was observed, large values $\nu\neq 0$ were found 
\cite{Badier:1981ti,Palestini:1985zc,Falciano:1986wk,Guanziroli:1987rp,Conway:1989fs}.
(However, in proton-proton and deuterium-proton collisions
\cite{Zhu:2006gx,Zhu:2008sj} the relation (\ref{Eq:Lam-Tung}) 
was found satisfied.)

An attractive explanation for the violation of the Lam-Tung 
relation (\ref{Eq:Lam-Tung}) is provided in the framework of 
TMDs \cite{Boer:1999mm} in terms of the Boer-Mulders function
$h_1^\perp$ \cite{Boer:1997nt}. In this approach  $\nu$ 
is related to $\sum_ae_a^2h_1^{\perp q}(x_1)h_1^{\perp \bar q}(x_2)$
\cite{Boer:1999mm}. 

It is of interest to learn about the Boer-Mulders function from DY and SIDIS 
because this T-odd TMD was predicted to have unusual universality properties.
On the basis of time-reversal arguments it was predicted \cite{Collins:2002kn} 
that $h_1^\perp$ in semi-inclusive deeply inelastic scattering (SIDIS) and 
in the Drell-Yan process (DY) have opposite signs,
\be\label{Eq:01}
     h_1^\perp(x,{\bf p}_T^2)_{DIS} =
    -h_1^\perp(x,{\bf p}_T^2)_{DY} \;.
\ee
Analogous relations are expected to hold also for the Sivers function
and other T-odd TMDs \cite{Collins:2002kn,Belitsky:2002sm}. 
The experimental check of such universality relations for T-odd TMDs
would provide a thorough test of our understanding of the factorization 
approach to transverse momentum dependent processes in terms of
$p_T$-dependent correlators 
\cite{Collins:2002kn,Belitsky:2002sm,Ji:2004wu,Collins:2004nx}.
Perspectives to test the universality relation
for the Sivers function where discussed in 
\cite{Efremov:2004tp,Collins:2005rq,Bianconi:2005yj,Anselmino:2009st,Kang:2009sm,
Bianconi:2006hc,Sissakian:2008th,Sissakian:2010zz,Bacchetta:2010si},
and for the Boer-Mulders function in 
\cite{Gamberg:2005ip,Bianconi:2006hc,Sissakian:2008th,Sissakian:2010zz,Bacchetta:2010si,
Sissakian:2005vd,Sissakian:2005yp}.

As we have seen in Sec.~\ref{Subsec-2E:Cahn-BM} the extraction of the 
Boer-Mulders function from SIDIS is hampered by substantial power-corrections
due to the Cahn-effect. It is interesting to ask whether the same difficulties
occur also in DY. In order to address this question a quantitative 
understanding of intrinsic $p_T$-effects in DY is necessary, 
and on the basis of the results from Section~\ref{Subsec-3B:sigma-qT-DY} 
we are prepared to have a look at that.

As in SIDIS, in the LO QCD (``tree level'') approach at low $q_T$
\cite{Arnold:2008kf} the coefficient of the $\cos\phi$ modulation in 
(\ref{Eq:sigma-azimuthal-DY}) is suppressed by one power of the 
large scale $Q$, but $\nu$ which is related to the Boer-Mulders effect 
is leading twist. And, as in  SIDIS, the Cahn effect
generates a $1/Q$-contribution to $\mu$ and a 
$1/Q^2$-power-correction to $\nu$. These contributions are
given by
\ba
      \mu_{\rm Cahn} &=& \;\,A\;\,\frac{q_T}{Q}\;\;\:
      \frac{\la p_{1T}^2\ra -\la p_{2T}^2\ra }{\la p_{1T}^2\ra +\la p_{2T}^2\ra }\;,\\
      \nu_{\rm Cahn} &=& A\;\frac{q_T^2}{Q^2}\;
      \biggl(\frac{\la p_{1T}^2\ra -\la p_{2T}^2\ra }{\la p_{1T}^2\ra +\la p_{2T}^2\ra }
      \biggr)^{\!2}\,,
\ea
where 
\be
     A = \Biggl[1+\frac{1}{2Q^2}
     \Biggl(
     \frac{4\la p_{1T}^2\ra \la p_{2T}^2\ra }{\la p_{1T}^2\ra +\la p_{2T}^2\ra }
     + \biggl(\frac{\la p_{1T}^2\ra -\la p_{2T}^2\ra }{\la p_{1T}^2\ra +\la p_{2T}^2\ra }\biggr)^{\!2}q_T^2
     \Biggr) \;\Biggr]^{-1} \;.
\ee
Notice that in both cases the effect vanishes, if the Gauss
widths are equal.
Which does not mean that if vanishes, if the colliding hadrons
are the same. For example, in proton-proton collisions the
Gauss widths of quarks and antiquarks enter, which could
be different. It is true that in Sec.~\ref{Sec-2:SIDIS} we saw no 
evidence for a flavor dependence of the Gauss widths, but we
should keep in mind that in SIDIS at Jefferson Lab and HERMES 
sea quarks do not play a dominant role.

\newpage

\begin{figure}[t!]
\begin{center}
\includegraphics[height=8cm]{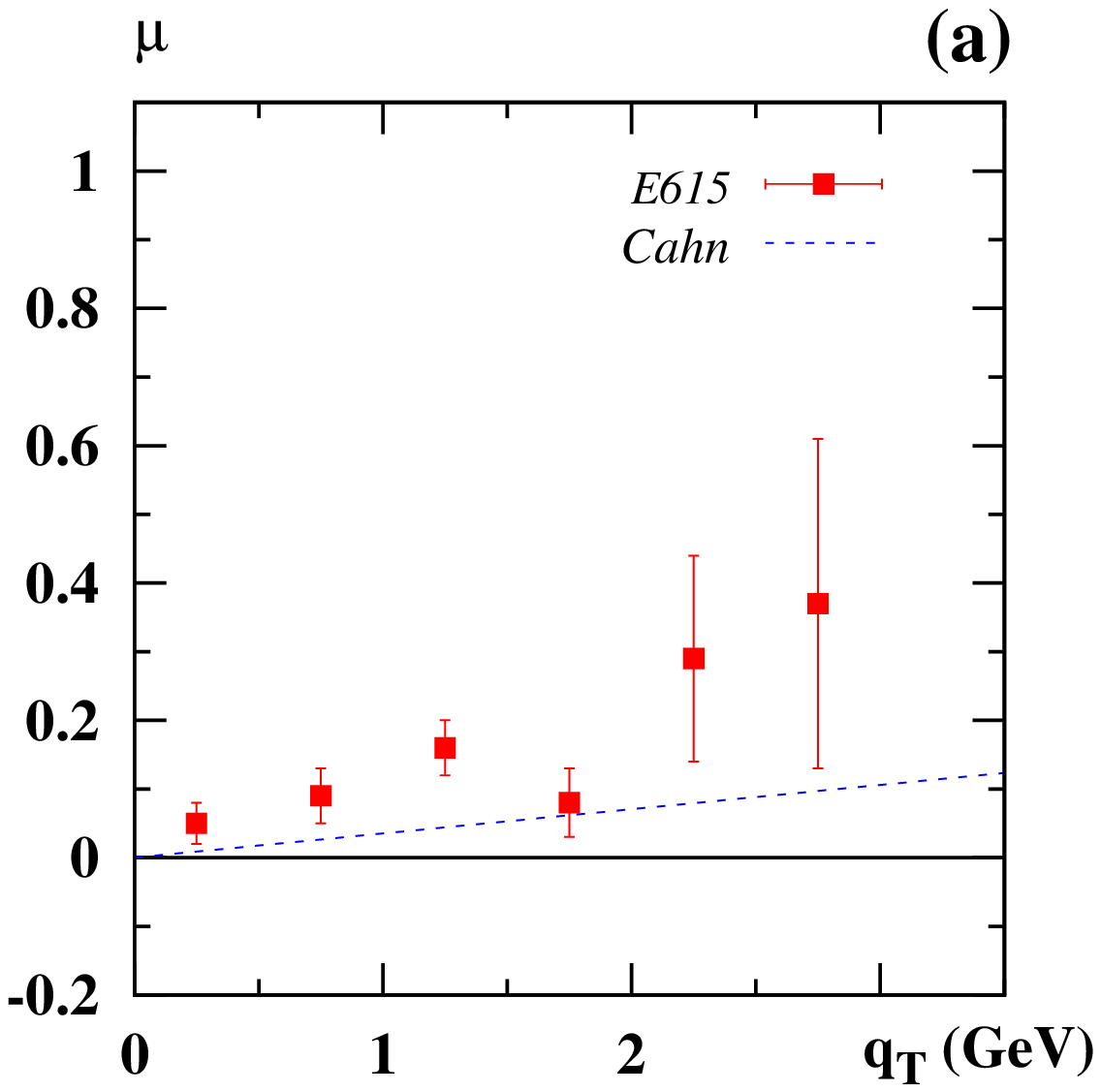}
\includegraphics[height=8cm]{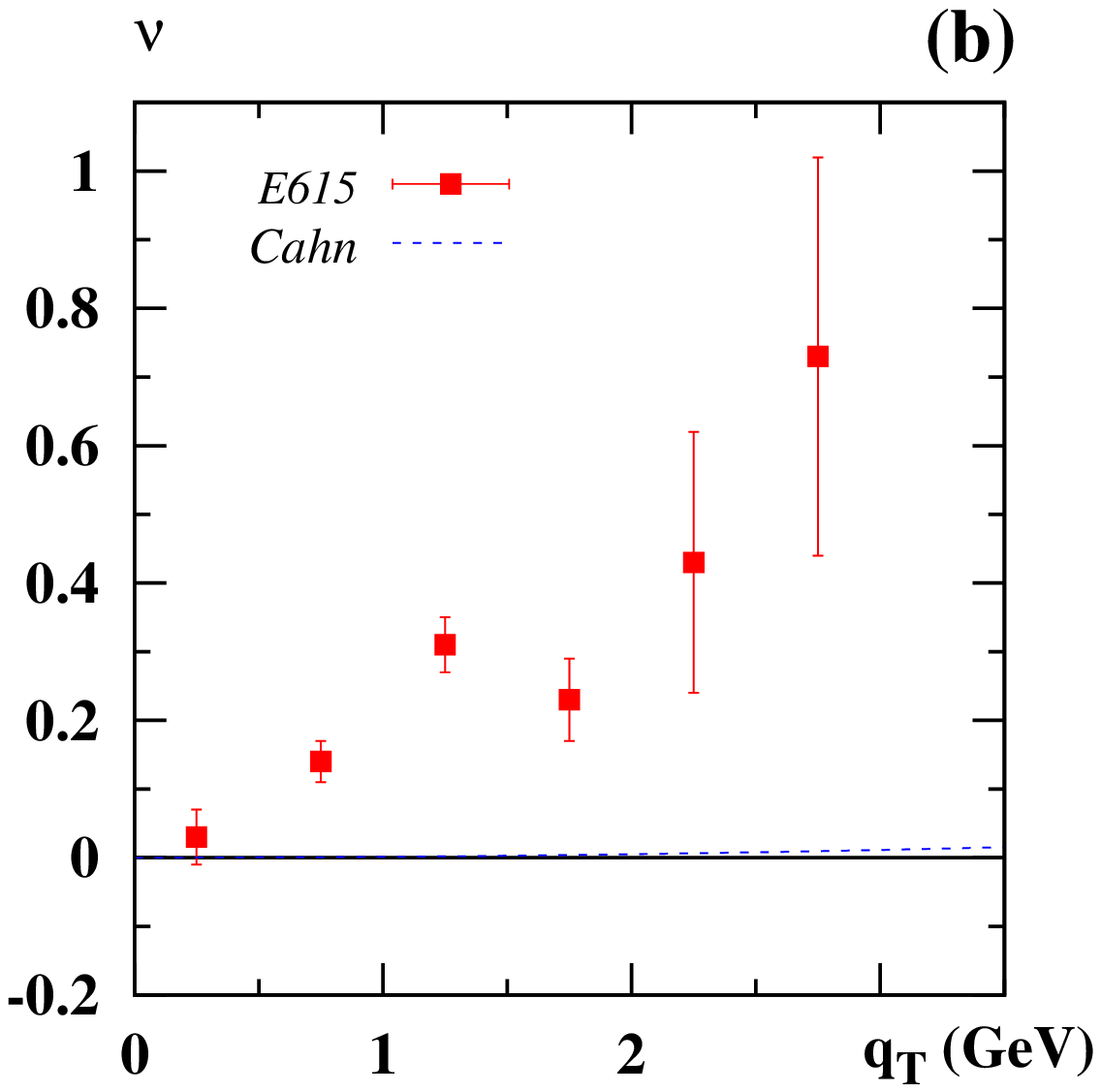}
\end{center}
        \caption{\label{Fig:Cahn-in-DY}
	The coefficient $\mu$ and $\nu$ in the azimuthal distribution
        (\ref{Eq:sigma-azimuthal-DY}) of the dileptons in DY as function
        of the dilepton transverse momentum $q_T$ in the
        Collins-Soper frame.
        Data are from the Fermilab E615 experiment \cite{Conway:1989fs}.
        The curves represent an estimate of the Cahn effect contribution 
        to these coefficients, see text.}
\end{figure}

With the results on the Gauss widths for pion and nucleon inferred
in Eq.~(\ref{DY-fix-widths}) we obtain the estimates for the contributions
of the Cahn effect to the coefficients $\mu$, $\nu$ shown in 
Fig.~\ref{Fig:Cahn-in-DY} in comparison to Fermilab E615 data
taken from $\pi^-$-nucleus collisions \cite{Conway:1989fs}.
According to the convention what is hadron 1 and hadron 2 
in \cite{Conway:1989fs} we have to identify
$\la p_{1T}^2\ra=\la p_{\pi T}^2\ra$ and 
$\la p_{2T}^2\ra=\la p_{\mbox{\tiny $N$}T}^2\ra$
in Eq.~(\ref{DY-fix-widths}).

%------ BEGIN FIGURE 9: cos(phi) AND cos(2phi) IN DY VS CAHN ---------- 
%-------- END FIGURE 9. ----------------------------------------------- 

We see in Fig.~\ref{Fig:Cahn-in-DY}a that the Cahn effect 
is able to account partly for the power-suppressed coefficient 
$\mu$ of the $\cos\phi$-modulation in (\ref{Eq:sigma-azimuthal-DY}).
At this point we should not be worried too much about the fact,
that in SIDIS at EMC the Cahn effect was able to account for 
the entire $\cos\phi$-modulation. Here we deal with a different
kinematics and TMDs replacing the role of fragmentation functions.
(For the latter reason one also cannot conclude anything on the
WW-type-approximations, see Sec.~\ref{Subsec-2D:Cahn-effect+EMC-data}.)
Also we have to keep in mind that the result for $\mu$ is strongly
sensitive to the difference of the Gauss widths for pion and
nucleon, which we estimated crudely in Eq.~(\ref{DY-fix-widths}).
In fact, if one was interested in that, one could describe
the $\mu$-coefficient entirely in terms of the Cahn effect
with the choice $\la p_{\pi T}^2\ra =1.3\,{\rm GeV}^2$ and 
$\la p_{\mbox{\tiny $N$}T}^2\ra =0.3\,{\rm GeV}^2$, but 
we shall refrain from doing this here.

What is important at this point is the observation, that
the Cahn-effect-power-correction to the $\nu$-coefficient
is negligible.
We conclude from this exercise that the present DY data provide
a ``safer'' way of accessing information on the Boer-Mulders
effect, in the sense that they are far less sensitive to
power-corrections as compared to available SIDIS data.
In practice, of course, data from both processes need to be explored
in order to test the universality prediction (\ref{Eq:01}).
Studies of the Boer-Mulders effect were presented in SIDIS
\cite{Oganesian:1997jq,Gamberg:2003ey,
Barone:2005kt,Barone:2006ws,Gamberg:2007wm,Zhang:2008ez,Barone:2008tn,Barone:2009hw}
and DY 
\cite{Gamberg:2005ip,Bianconi:2006hc,Sissakian:2008th,Sissakian:2010zz,Bacchetta:2010si,
Sissakian:2005vd,Sissakian:2005yp,Lu:2009ip},
see also \cite{Zhou:2009rp}.

\newpage
%=== SECTION 4: s-DEPENDENCE =========================================
\section{Energy dependence of intrinsic transverse momenta 
in DY and SIDIS}
\label{Sec-4:s-dependence}

In Eqs.~(\ref{Eq:DY-qT2-in-p-p},~\ref{Eq:DY-qT2-in-p-pi}) we were able to 
combine  information on the $q_T$-dependence of DY cross sections from 
different experiments to arrive at (\ref{DY-fix-widths}), because of comparable
kinematics. In general experiments performed at different energies have 
to be compared with care.
In DY the mean lepton momentum square $\la q_T^2\ra$ is energy ($s$-)dependent. 
This dependence is different for $i=\pi N$ or $p N$ induced Drell-Yan.
For $50\,{\rm GeV}^2 < s < 600\,{\rm GeV}^2$ it can roughly be described as 
\cite{Malhotra:1982cx}
\ba
	\la q_T^2(s)\ra_i = A_i + B_i\,s \;,\;\;\;\;\;
        A_{\pi N}=(0.59\pm 0.05)\,{\rm GeV}^2,\;\;
        B_{\pi N}=(2.8 \pm 0.2 )\cdot 10^{-3},\;\;\nonumber\\
        A_{pN}=(0.52\pm 0.11)\,{\rm GeV}^2,\;\;
        \;B_{pN}=(1.4 \pm 0.2 )\cdot 10^{-3}.\;\;\label{Eq:qT2-vs-s-fit}
\ea
%------ BEGIN FIGURE 10+11: MALHOTRA & Phperp^2 VS s IN SIDIS ----------
\begin{wrapfigure}[36]{RD}{8cm}
        \vspace{-0.8cm}
	\includegraphics[height=6.5cm]{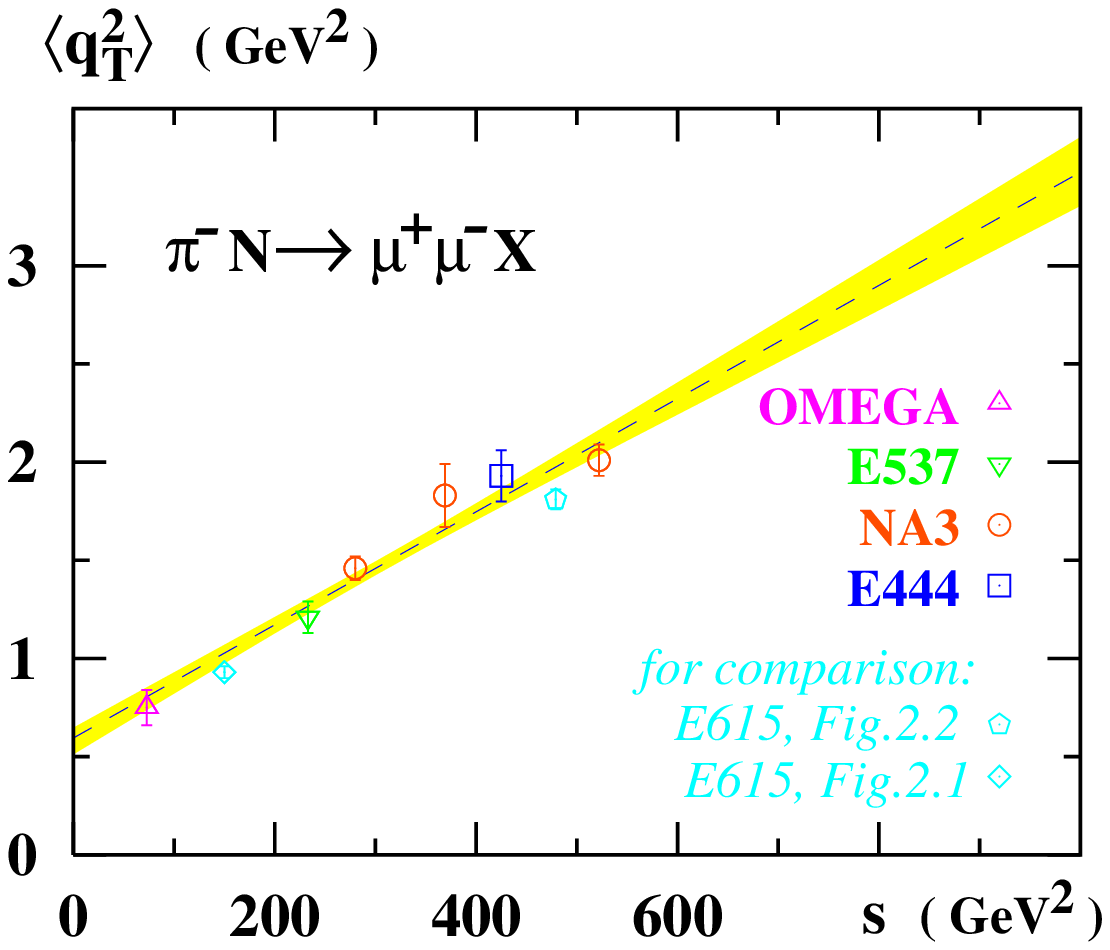}
        \vspace{-0.2cm} 
        \caption{\label{Fig03:pT2-vs-s}
	Mean dimuon transverse momentum square $\la q_T^2\ra$ as function of
	the center of mass energy square, $s$, 
	in $\pi^-N$ induced Drell-Yan. Following \cite{Malhotra:1982cx}.}

        \vspace{0.5cm}

	\includegraphics[width=7cm]{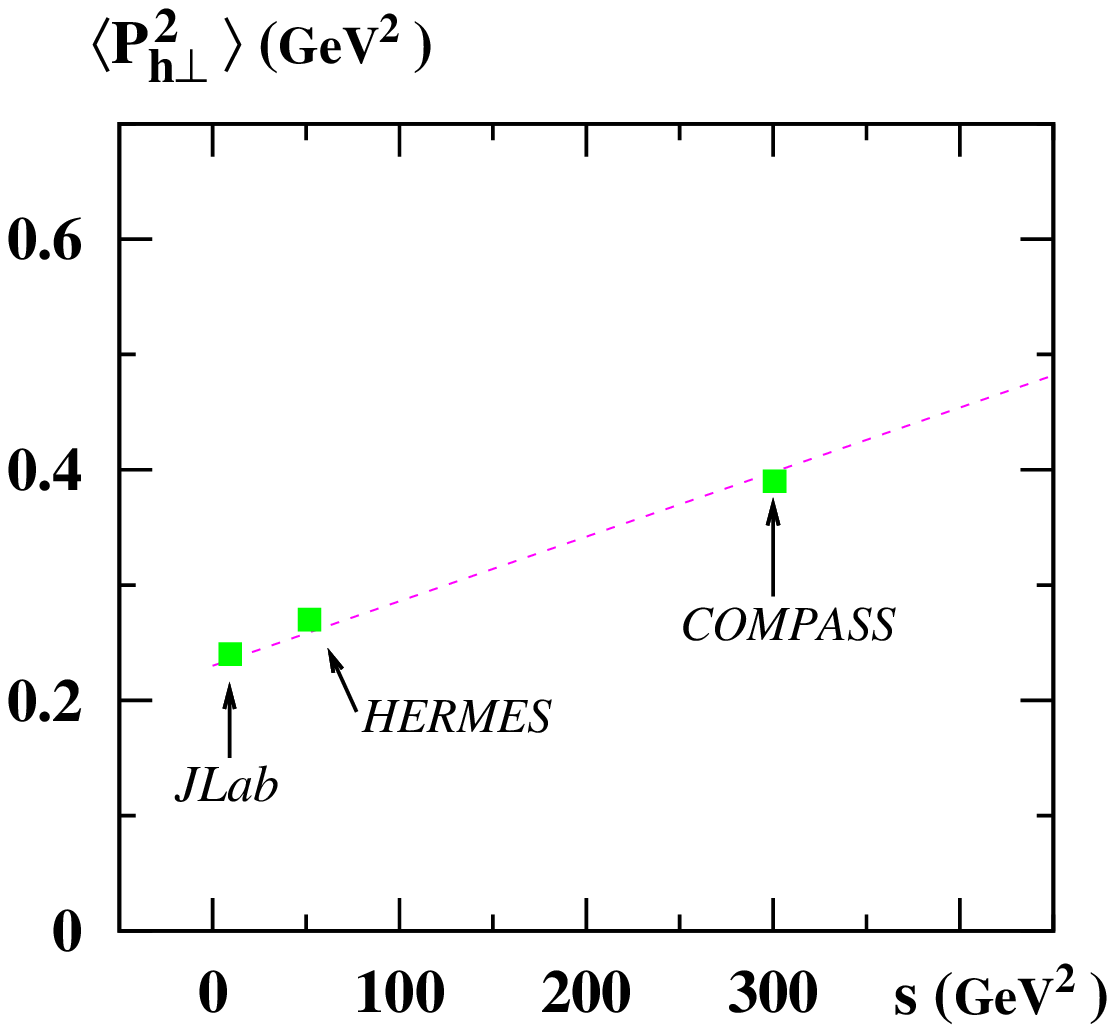}
        \vspace{-0.2cm} 
        \caption{\label{Fig10:pT2-vs-s-SIDIS}
        Mean square transverse momenta $\la P^2_{h\perp}(z)\ra$ 
        in SIDIS around $z\sim0.5$ as function of $s$ from
        Jefferson Lab \cite{Osipenko,Mkrtchyan:2007sr},
        HERMES \cite{Airapetian:2009jy}, COMPASS 
        \cite{Ageev:2006da}.}

\end{wrapfigure}
%------ END FIGURE 10+11 ---------------------------------------------

This is just one way of parametrizing the $s$-dependence.
The data are compatible also with a linear in $\sqrt{s}$
increase of $\la q_T^2\ra$ \cite{Malhotra:1982cx},
while in QCD one would rather expect a logarithmic increase. 
But in a limited $s$-range an effective parametrization of 
the type (\ref{Eq:qT2-vs-s-fit}) works reasonably well.
In any case, $\la q_T^2\ra$ increases with energy, which reflects the
transverse momentum broadening due to gluon radiation \cite{Collins:1984kg}.

Fig.~\ref{Fig03:pT2-vs-s} shows $\la q_T^2\ra$ in $\pi N$ induced reactions
and the original fit from \cite{Malhotra:1982cx}.
The values of  $\la q_T^2\ra_{\pi N}$ deduced from Figs.~\ref{FigUV:DY-qT-qT2} 
and \ref{FigXY:DY-qT} are shown for comparison but not included in the fits.
Assuming the Gauss model (\ref{Eq:DY-qT-01a}) we obtain  from 
(\ref{Eq:qT2-vs-s-fit}) the following effective $s$-dependence of 
the intrinsic transverse parton momenta in the hadron $h$:
\ba\label{Eq:eff-s-dependence-0}
	\la p_T^2(s)\ra_h &\approx& \la p_T^2(0)\ra + C_h\;s
        \phantom{\biggl|}\\
\label{Eq:eff-s-dependence-1}
&&	\la p_T^2(0)\ra = 0.3\,{\rm GeV}^2\\
   \label{Eq:eff-s-dependence-2}
&&	C_h = 10^{-3}\times\cases{2.1 & for $h=\pi$, \cr
				  0.7 & for $h= p$.} 
	\label{Eq:pT2-vs-s-approx}\ea

At this point two questions arise.
First, are transverse parton momenta in DY and SIDIS
compatible? Second, if so, is there any indication of 
transverse momentum broadening in SIDIS too?

Concerning the first question, it is encouraging to 
observe that (\ref{Eq:eff-s-dependence-0}) ``predicts'' 
for HERMES ($s=52\,{\rm GeV}^2$)
the result $\la p_T^2(s)\ra|_{\rm HERMES}=0.34\,{\rm GeV}^2$
which is within the error bars of the Gauss width in
(\ref{Eq:new-fit-HERMES}). 
Probably it would be more consistent to use in SIDIS the 
photon-hadron center of mass energy square $W^2$ instead of
$s$.
However, in view of the uncertainties in (\ref{Eq:new-fit-HERMES})
and (\ref{Eq:eff-s-dependence-0}---\ref{Eq:eff-s-dependence-2})
this is numerically of little relevance.

Concerning the second question, let us compare the mean
square transverse momenta $\la P^2_{h\perp}(z)\ra$ in SIDIS 
from various experiments at a common value of $z$, 
let us say $0.5<z<0.6$ (as different ranges are covered
we cannot compare averages over $z$).
For Jefferson Lab we use 
$\la P_{h\perp}^2(z)\ra=0.24\,{\rm GeV}^2$ at $z=0.55$ 
from CLAS \cite{Osipenko}, which describes well the Hall C 
data \cite{Mkrtchyan:2007sr}, see Fig.~\ref{Fig-03:Hall-C}.
For HERMES we take $\la P_{h\perp}^2(z)\ra=0.27\,{\rm GeV}^2$ at $z=0.52$
\cite{Airapetian:2009jy}. For COMPASS we use
 $\la P_{h\perp}(z)\ra=0.55\,{\rm GeV}$ at $z=0.566$
from \cite{Ageev:2006da} which we convert by means of
(\ref{Eq:Gauss-predict}).

Fig.~\ref{Fig10:pT2-vs-s-SIDIS} must be interpreted with care.
The shown $\la P_{h\perp}^2(z)\ra$ were obtained in 
different ways, span a small $s$-range, and have 
systematic uncertainties except for the HERMES value.
For a conclusive comparison acceptance corrected data
are needed from all experiments. 
Nevertheless, we see a tendency for an increase in $s$ with a 
slope which is $20\,\%$ lower than (\ref{Eq:eff-s-dependence-2}),
see Fig.~\ref{Fig10:pT2-vs-s-SIDIS}. We made no effort to estimate 
the uncertainty of (\ref{Eq:eff-s-dependence-2}) but it is presumably
not smaller than $20\,\%$. So the $s$-slopes in SIDIS and DY are
compatible. 

\newpage

Notice that in SIDIS also the ``broadening'' of the width of the 
transverse momenta in the fragmentation function contributes, and 
we above tacitly assumed that both $\la p_T^2\ra$ in $f_1^a(x,p_T)$
and $\la K_T^2\ra$ in $D_1^a(z,K_T)$ increase with $s$ at a similar rate. 
In fact, would we extrapolate from HERMES at $s=52\,{\rm GeV}^2$,
Eq.~(\ref{Eq:new-fit-HERMES}), assuming that $\la p_T^2\ra$ 
increases with $s$ at the rate (\ref{Eq:eff-s-dependence-2}) but 
keeping $\la K_T^2\ra$ at its initial value in (\ref{Eq:new-fit-HERMES}), 
we would ``predict'' $\la P_{h\perp}^2(z)\ra\sim (0.30\pm0.03)\,{\rm GeV}^2$ 
at $z=0.566$ for COMPASS which strongly underestimates the measured value 
$\la P_{h\perp}^2(z)\ra \simeq 0.39\,{\rm GeV}^2$ \cite{Ageev:2006da}.
Thus, the result in Fig.~\ref{Fig10:pT2-vs-s-SIDIS} also indicates 
$\la K_T^2\ra$-broadening in the fragmentation function.

Finally, let us comment on EMC data on $A_{UU}^{\cos\phi}$.
In Sec.~\ref{Subsec-2D:Cahn-effect+EMC-data} we have seen that the 
Cahn effect can explain these data with the Gauss widths from HERMES. 
At first glance, this seems to imply that the Gauss widths and EMC 
(where $s=525\,{\rm GeV}^2$) are as large as at HERMES, while 
Fig.~\ref{Fig10:pT2-vs-s-SIDIS} would suggest that they should 
be substantially larger.
Indeed, if we extrapolate from Fig.~\ref{Fig10:pT2-vs-s-SIDIS} to EMC 
energies then we obtain for EMC about 2 times larger widths resulting in an 
about (30--40)$\,\%$ larger Cahn contribution to the $\cos\phi$-modulation.
At this point, however, we have to recall that the Cahn effect description
of this observable requires the neglect of quark-gluon-correlations
in (\ref{Eq:FUUcosphi}). Such quark-gluon-correlations could be as large 
as (30--40)$\,\%$ compared to the contributions from quark correlators,
as was found in \cite{Accardi:2009au} in the context of a different 
observable.
Thus, the intrinsic transverse momenta at EMC could be well compatible
with the picture in Fig.~\ref{Fig10:pT2-vs-s-SIDIS}. Future data from 
HERMES, COMPASS, and Jefferson Lab on  $A_{UU}^{\cos\phi}$ will clarify 
the situation.
 
To summarize: we conclude that the non-perturbative mechanisms responsible 
for intrinsic transverse parton momenta in DY and SIDIS are apparently 
compatible. On the basis of TMD-factorization and universality 
this is expec\-ted, but in view of the typically different energy scales 
probed in these processes it is not straight-forward to~see.

%newpage
%=== SECTION 5: s-DEPENDENCE =========================================
\section{Conclusions}
\label{Sec-5:conclusions}

In this work we discussed and reviewed the present understanding 
of intrinsic transverse parton momenta in SIDIS and DY.
An important aim was to demonstrate that the popular Gauss model 
works very well in these processes. 
A similar in spirit study was presented in \cite{D'Alesio:2004up}.
But meanwhile many more data especially from SIDIS emerged,
which allow more conclusive tests of the Gauss model, and make an update 
of previous results \cite{D'Alesio:2004up,Collins:2005ie,Anselmino:2005nn} 
possible.

More precisely, in the case of DY we have seen that the 
Gauss model is applicable if the transverse (dilepton)
momenta much smaller than the hard scale. 
In SIDIS at energies probed at Jefferson Lab or HERMES we
have seen that the Gauss model well describes all available
data on cross sections or transverse hadron momenta. 
It is interesting to remark that the Gauss model
has received certain support from models \cite{Avakian:2010br,Boffi:2009sh} 
and lattice QCD \cite{lattice-TMD}.

We discussed the known fact that the Gauss widths increase
with energy in DY. By comparing transverse hadron momenta at 
Jefferson Lab, HERMES and COMPASS we found indications for the energy 
dependence of the Gauss widths in SIDIS too.
More precise data from SIDIS are needed, but on the 
basis of what is available now, we conclude that the
intrinsic transverse parton momenta in SIDIS and DY
are compatible. This demonstrates the universality of
the $p_T$-dependence of $f_1^a(x,p_T)$ in SIDIS and DY
and supports the factorization approach in terms of 
unintegrated correlators, although the support has
presently a qualitative character.

Our results are of importance for many practical applications.
First, the Gauss model can now be used in SIDIS as an effective tool for 
the description of transverse parton momenta with more confidence.
On the basis of presently available data we have seen no evidence
for a worthwhile mentioning flavor- or $x$- or $z$-dependence of 
the Gauss widths.
Second, we learned that the Gauss widths are energy dependent also
in SIDIS, which is of importance when quantitatively comparing results 
from Jefferson Lab, HERMES and COMPASS. The energy dependence of the Gauss widths
in DY was well known before \cite{Malhotra:1982cx}.
Third, a solid understanding of $p_T$-effects and their energy dependence
is indispensable in order to study the azimuthal and spin asymmetries
in DY and SIDIS.  
Our results will in particular be helpful to make estimates for 
the planned or proposed DY physics programs at COMPASS, GSI, 
J-PARC, U-70 where the process will be probed at different energies.

Especially in the context of the azimuthal asymmetries in unpolarized
SIDIS a good quantitative understanding of intrinsic transverse momenta
is of importance in the context of the Boer-Mulders effect. We have estimated
that the associated $\cos(2\phi)$--modulation of the SIDIS cross section
receives at Jefferson Lab, HERMES, and COMPASS energies sizeable $1/Q^2$ power 
corrections from the Cahn effect --- which is just one of the possible 
non-factorizing power corrections --- in agreement with results
from other studies \cite{Barone:2008tn,Barone:2009hw}. Due to the larger $Q$
in the available DY data the Cahn effect does not hamper the extraction 
of the Boer-Mulders function.

As a byproduct, we confirmed the observation \cite{Anselmino:2005nn} that the 
Cahn effect can account for the EMC data \cite{Aubert:1983cz,Arneodo:1986cf} 
on the ``twist-3'' $\cos\phi$--modulation in the unpolarized SIDIS cross 
section, using the Gauss model parameters from HERMES. However, we were 
lead to the suspicion that this good description is likely to be due
to a cancellation of effects due to the Gauss width broadening
at the larger EMC energies and terms neglected in the so-called
Wandzura-Wilczek-type approximation needed to justify here the 
``Cahn-effect-only'' approximation.

To conclude, the Gauss model with carefully taken into account 
energy dependence of the Gauss widths is not only a convenient
but also within a good accuracy well justified tool to describe
intrinsic transverse parton momenta. This approach, especially
now at the early state of art of studies of azimuthal and spin 
asymmetries in SIDIS and DY, represents a sufficient approximation for 
many practical purposes. The Gauss Ansatz remains to be tested,
in particular, when polarization phenomena are included, and
future data may demand to refine it, which will improve 
our understanding of intrinsic transverse parton momenta.

Future steps, necessary when one will be interested in high 
precision and/or in going to high energies, for example DY at RHIC 
with $\sqrt{s}=200\,{\rm GeV}$, will include the treatment within
the Collins-Soper-Sterman formalism \cite{Collins:1984kg}, 
as implemented for instance in \cite{Landry:2002ix} and consideration 
of scale dependence \cite{Idilbi:2004vb,Cherednikov:2007tw}.

\ \\
{\bf Acknowledgements.}
We are grateful to H.~Avakian, A.~V.~Efremov, F.~Giordano, R.~Lamb, 
M.~Osipenko, G.~Schnell for discussions and/or providing final or
preliminary data. This work is partially supported by the Verbundforschung 
``Hadronen und Kerne'' of the BMBF. 
A.~M.~acknowledges the support of the NSF under Grant No.~PHY-0855501.
P.~S.~is supported by DOE contract No. DE-AC05-06OR23177, under which Jefferson
Science Associates, LLC operates Jefferson Lab. 
T.~T.~is supported by the Cusanuswerk.

%===================  REFERENCES =====================================

% ** ** **  ** ** **  ** ** **  ** ** **  ** ** **  ** ** **  ** ** ** 
% HOW TO ORDER THE REFERENCES IN A DRAFT:
% send mail  To: slaclib2@slac.stanford.edu
% with  Subject: generate eu
% with the body: latex-file of draft with Refs. in spires-format  
% then an ordered, and updated list will be sent back!  :)
% ** ** **  ** ** **  ** ** **  ** ** **  ** ** **  ** ** **  ** ** ** 


\begin{thebibliography}{999}

\bibitem{Collins:1981uw}
   J.~C.~Collins and D.~E.~Soper,
   %``Parton Distribution And Decay Functions,''
   Nucl.\ Phys.\  B {\bf 194}, 445 (1982).
   %%CITATION = NUPHA,B194,445;%%

\bibitem{Collins:1981uk}
  J.~C.~Collins and D.~E. Soper,
  % ``BACK-TO-BACK JETS IN QCD,''
  Nucl.\ Phys.\ B {\bf 193}, 381 (1981)
  [Erratum-ibid.\ B {\bf 213}, 545 (1983)].
  %%CITATION = NUPHA,B193,381;%%

\bibitem{Collins:1999dz}
  J.~C.~Collins, F.~Hautmann,
  % ``Infrared divergences and non-lightlike eikonal lines in Sudakov
  %processes,''
  Phys.\ Lett.\  B {\bf 472} (2000) 129,
  %[arXiv:hep-ph/9908467].
  %%CITATION = PHLTA,B472,129;%%
%\bibitem{Collins:2000gd}
  % J.~C.~Collins and F.~Hautmann,
  % ``Soft gluons and gauge-invariant subtractions in NLO parton-shower Monte
  %Carlo event generators,''
  JHEP {\bf 0103} (2001) 016. 
  %[arXiv:hep-ph/0009286].
  %%CITATION = JHEPA,0103,016;%%

\bibitem{Collins:2003fm}
  J.~C.~Collins,
  %``What exactly is a parton density?,''
  Acta Phys.\ Polon.\  B {\bf 34} (2003) 3103
  [arXiv:hep-ph/0304122].
  %%CITATION = APPOA,B34,3103;%%

\bibitem{Collins:2007ph}
  J.~C.~Collins, T.~C.~Rogers and A.~M.~Stasto,
  %``Fully Unintegrated Parton Correlation Functions and Factorization in Lowest
  %Order Hard Scattering,''
  Phys.\ Rev.\  D {\bf 77} (2008) 085009.
  %[arXiv:0708.2833 [hep-ph]].
  %%CITATION = PHRVA,D77,085009;%%

\bibitem{Hautmann:2007uw}
  F.~Hautmann,
  %``Endpoint singularities in unintegrated parton distributions,''
  Phys.\ Lett.\  B {\bf 655} (2007) 26. 
  %[arXiv:hep-ph/0702196].
  %%CITATION = PHLTA,B655,26;%%

\bibitem{Collins:2008ht}
   J.~Collins,
   %``Rapidity divergences and valid definitions of parton densities,''
   PoS {\bf LC2008}, 028 (2008)
   [arXiv:0808.2665 [hep-ph]].
   %%CITATION = POSCI,LC2008,028;%%

\bibitem{Georgi:1977tv}
  H.~Georgi and H.~D.~Politzer,
  %``Clean Tests Of QCD In Mu P Scattering,''
  Phys.\ Rev.\ Lett.\  {\bf 40} (1978) 3.
  %%CITATION = PRLTA,40,3;%%

\bibitem{Cahn:1978se}
  R.~N.~Cahn,
  %``Azimuthal Dependence In Leptoproduction: A Simple Parton Model
  %Calculation,''
  Phys.\ Lett.\  B {\bf 78} (1978) 269.
  %%CITATION = PHLTA,B78,269;%%

\bibitem{Konig:1982uk}
  A.~Konig and P.~Kroll,
  %``A Realistic Calculation Of The Azimuthal Asymmetry In Semiinclusive Deep
  %Inelastic Scattering,''
  Z.\ Phys.\  C {\bf 16} (1982) 89.
  %%CITATION = ZEPYA,C16,89;%%

\bibitem{Efremov:1980kz}
  A.~V.~Efremov and A.~V.~Radyushkin,
  %``Field Theory Approach To Processes With Large Momentum Transfer.
  % Ii. Production Of Massive Lepton Pairs,''
  JINR E2-11726 (1978), 
  Theor.\ Math.\ Phys.\  {\bf 44}, 664 (1981)
  [Teor.\ Mat.\ Fiz.\  {\bf 44}, 157 (1980)].
  %%CITATION = TMFZA,44,157;%%

\bibitem{Collins:1984kg}
  J.~C.~Collins, D.~E.~Soper and G.~Sterman,
  %``Transverse Momentum Distribution In Drell-Yan Pair And W And Z Boson
  %Production,''
  Nucl.\ Phys.\  B {\bf 250} (1985) 199.
  %%CITATION = NUPHA,B250,199;%%

\bibitem{Sivers:1989cc}
  D.~W.~Sivers,
  %``Single Spin Production Asymmetries From The Hard Scattering Of Point-Like
  %Constituents,''
  Phys.\ Rev.\ D {\bf 41}, 83 (1990),
  %%CITATION = PHRVA,D41,83;%%
%\bibitem{Sivers:1990fh}
  % D.~W.~Sivers,
  %``Hard Scattering Scaling Laws For Single Spin Production Asymmetries,''
  Phys.\ Rev.\ D {\bf 43}, 261 (1991).
  %%CITATION = PHRVA,D43,261;%%

\bibitem{Efremov:1992pe}
  A.~V.~Efremov, L.~Mankiewicz and N.~A.~Tornqvist,
  %``Jet handedness as a measure of quark and gluon polarization,''
  Phys.\ Lett.\ B {\bf 284} (1992) 394.
  %%CITATION = PHLTA,B284,394;%%
  %%Cited 74 times in SPIRES-HEP

\bibitem{Collins:1992kk}
  J.~C.~Collins,
  % Fragmentation of transversely polarized quarks probed in transverse
  % momentum distributions,
  Nucl.\ Phys.\ B {\bf 396}, 161 (1993) [arXiv:hep-ph/9208213].
  %%CITATION = HEP-PH 9208213;%%

\bibitem{Collins:1993kq}
  J.~C.~Collins, S.~F.~Heppelmann and G.~A.~Ladinsky,
  %``Measuring transversity densities in singly polarized hadron hadron and
  %lepton - hadron collisions,''
  Nucl.\ Phys.\ B {\bf 420} (1994) 565
  [arXiv:hep-ph/9305309].
  %%CITATION = HEP-PH 9305309;%%

\bibitem{Kotzinian:1994dv}
  A.~Kotzinian,
  %``New quark distributions and semiinclusive electroproduction on the
  %polarized nucleons,''
  Nucl.\ Phys.\  B {\bf 441} (1995) 234
  [arXiv:hep-ph/9412283].
  %%CITATION = NUPHA,B441,234;%%

\bibitem{Mulders:1995dh}
  P.~J.~Mulders and R.~D.~Tangerman,
  %``The complete tree-level result up to order 1/Q for polarized
  %deep-inelastic leptoproduction,''
  Nucl.\ Phys.\ B {\bf 461} (1996) 197
  % [Erratum-ibid.\ B {\bf 484} (1997) 538]
  and {\bf 484} (1997) 538E
  [arXiv:hep-ph/9510301].
  %%CITATION = HEP-PH 9510301;%%

\bibitem{Boer:1997nt}
  D.~Boer and P.~J.~Mulders,
  %``Time-reversal odd distribution functions in leptoproduction,''
  Phys.\ Rev.\ D {\bf 57}, 5780 (1998)
  [arXiv:hep-ph/9711485].
  %%CITATION = HEP-PH 9711485;%%

\bibitem{Boer:1999mm}
  D.~Boer,
  %``Investigating the origins of transverse spin asymmetries at RHIC,''
  Phys.\ Rev.\ D {\bf 60}, 014012 (1999)
  [arXiv:hep-ph/9902255].
  %%CITATION = HEP-PH 9902255;%%

\bibitem{Bacchetta:1999kz}
  A.~Bacchetta, M.~Boglione, A.~Henneman and P.~J.~Mulders,
  %``Bounds on transverse momentum dependent distribution and fragmentation
  %functions,''
  Phys.\ Rev.\ Lett.\  {\bf 85}, 712 (2000)
  [arXiv:hep-ph/9912490].
  %%CITATION = PRLTA,85,712;%% %%CITATION = PHLTA,B387,577;%%

\bibitem{Boer:1997mf}
  D.~Boer, R.~Jakob and P.~J.~Mulders,
  %``Asymmetries in polarized hadron production in e+ e- annihilation up to
  %order 1/Q,''
  Nucl.\ Phys.\ B {\bf 504} (1997) 345
  [arXiv:hep-ph/9702281].
  %%CITATION = HEP-PH 9702281;%%

\bibitem{Brodsky:2002cx}
  S.~J.~Brodsky, D.~S.~Hwang and I.~Schmidt,
  % Final-state interactions and single-spin asymmetries in semi-inclusive
  % deep inelastic scattering,
  Phys.\ Lett.\ B {\bf 530}, 99 (2002)
  [arXiv:hep-ph/0201296]; %%CITATION = HEP-PH 0201296;%%
%\bibitem{Brodsky:2002rv}
  % S.~J.~Brodsky, D.~S.~Hwang and I.~Schmidt,
  % ``Initial-state interactions and single-spin asymmetries in Drell-Yan
  % processes,''
  Nucl.\ Phys.\ B {\bf 642}, 344 (2002)
  [arXiv:hep-ph/0206259].
  %%CITATION = HEP-PH 0206259;%%

\bibitem{Collins:2002kn}
  J.~C.~Collins,
  % Leading-twist single-transverse-spin asymmetries: Drell-Yan and
  % deep-inelastic scattering,
  Phys.\ Lett.\ B {\bf 536}, 43 (2002) [arXiv:hep-ph/0204004].
  %%CITATION = HEP-PH 0204004;%%

\bibitem{Belitsky:2002sm}
  A.~V.~Belitsky, X.~Ji and F.~Yuan,
  % ``Final state interactions and gauge invariant parton distributions,''
  Nucl.\ Phys.\ B {\bf 656}, 165 (2003)
  [arXiv:hep-ph/0208038].
  %%CITATION = HEP-PH 0208038;%%
%\bibitem{Ji:2002aa}
  X.~D.~Ji and F.~Yuan,
  % ``Parton distributions in light-cone gauge:
  % Where are the final-state  interactions?,''
  Phys.\ Lett.\ B {\bf 543}, 66 (2002)
  [arXiv:hep-ph/0206057].
  %%CITATION = HEP-PH 0206057;%%
  \\
%\bibitem{Boer:2003cm}
  D.~Boer, P.~J.~Mulders and F.~Pijlman,
  %``Universality of T-odd effects in single spin and azimuthal asymmetries,''
  Nucl.\ Phys.\ B {\bf 667}, 201 (2003)
  [arXiv:hep-ph/0303034].
  %%CITATION = HEP-PH 0303034;%%

\bibitem{Ji:2004wu}
  X.~D.~Ji, J.~P.~Ma and F.~Yuan,
  %``QCD factorization for semi-inclusive deep-inelastic scattering at low
  %transverse momentum,''
  Phys.\ Rev.\ D {\bf 71}, 034005 (2005)
  [arXiv:hep-ph/0404183],
  %%CITATION = HEP-PH 0404183;%%
%\bibitem{Ji:2004xq}
  % X.~d.~M.~Ji, J.~P.~M.~Ma and F.~Yuan,
  %``QCD factorization for spin-dependent cross sections in DIS and Drell-Yan
  %processes at low transverse momentum,''
  Phys.\ Lett.\ B {\bf 597}, 299 (2004)
  [arXiv:hep-ph/0405085].
   %%CITATION = HEP-PH 0405085;%%

\bibitem{Collins:2004nx}
  J.~C.~Collins and A.~Metz,
  %``Universality of soft and collinear factors in hard-scattering
  %factorization,''
  Phys.\ Rev.\ Lett.\  {\bf 93}, 252001 (2004)
  [arXiv:hep-ph/0408249].
  %%CITATION = HEP-PH 0408249;%%

\bibitem{Bunce:1976yb}
  G.~Bunce {\it et al.},
  %``Lambda0 Hyperon Polarization In Inclusive Production By 300-GeV Protons On
  %Beryllium,''
  Phys.\ Rev.\ Lett.\  {\bf 36} 1113, (1976).
  %%CITATION = PRLTA,36,1113;%%

\bibitem{Adams:1991rw}
  D.~L.~Adams {\it et al.}, % [E581 Collaboration],
  %``Comparison of spin asymmetries and cross-sections in pi0 production by
  %200-GeV polarized anti-protons and protons,''
  Phys.\ Lett.\ B {\bf 261}, 201 and {\bf 264}, 462 (1991),
  %%CITATION = PHLTA,B261,201;%%
%\bibitem{Adams:1991cs}
  % D.~L.~Adams {\it et al.}  [FNAL-E704 Collaboration],
  %``Analyzing power in inclusive pi+ and pi- production at high x(F) with a
  %200-GeV polarized proton beam,''
  Phys.\ Lett.\ B {\bf 264}, 462 (1991),
  %%CITATION = PHLTA,B264,462;%%
%\bibitem{Adams:1991ru}
  % D.~L.~Adams {\it et al.}  [E581 Collaboration],
  %``Large x_F spin asymmetry in pi0 production by 200-GeV polarized protons,''
  Z.\ Phys.\ C {\bf 56}, 181 (1992).
  %%CITATION = ZEPYA,C56,181;%%

\bibitem{Adams:2003fx}
  J.~Adams {\it et al.}  [STAR Collaboration],
  %``Cross sections and transverse single-spin asymmetries in forward neutral
  %pion production from proton collisions at s**(1/2) = 200-GeV,''
  Phys.\ Rev.\ Lett.\  {\bf 92}, 171801 (2004)
  [arXiv:hep-ex/0310058].\\
  %%CITATION = HEP-EX 0310058;%%
%\bibitem{Adler:2005in}
  S.~S.~Adler {\it et al.}  [PHENIX Collaboration],
  %``Measurement of transverse single-spin asymmetries for mid-rapidity
  %production of neutral pions and charged hadrons in polarized p + p
  %collisions at s**(1/2) = 200-GeV,''
  Phys.\ Rev.\ Lett.\  {\bf 95} (2005) 202001
  [arXiv:hep-ex/0507073].\\
  %%CITATION = PRLTA,95,202001;%%
%\bibitem{Videbaek:2005fm}
  F.~Videbaek  [BRAHMS Collaboration],
  %``Single spin asymmetries in the BRAHMS experiment,''
  AIP Conf.\ Proc.\  {\bf 792} (2005) 993
  [arXiv:nucl-ex/0508015].\\
  %%CITATION = APCPC,792,993;%%
%\bibitem{Aidala:2006eq}
  C.~A.~Aidala,
  %``Measurement of the transverse single-spin asymmetry for mid-rapidity
  %production of neutral pions in polarized p + p collisions at 200-GeV
  %center-of-mass energy,''
  arXiv:hep-ex/0601009.\\
  %%CITATION = HEP-EX/0601009;%
%\bibitem{Arsene:2008mi}
  I.~Arsene {\it et al.}  [BRAHMS Collaboration],
  %``Single Transverse Spin Asymmetries of Identified Charged Hadrons in
  %Polarized p+p Collisions at $\sqrt{s}$ = 62.4 GeV,''
  Phys.\ Rev.\ Lett.\  {\bf 101}, 042001 (2008)
  [arXiv:0801.1078 [nucl-ex]].
  %%CITATION = PRLTA,101,042001;%%

%====== DATA ON AZIMUTHAL ASYMMETRIES IN UNPOL. DY
\bibitem{Aubert:1983cz}
  J.~J.~Aubert {\it et al.}  [European Muon Collaboration],
  %``Measurement of hadronic azimuthal distributions in deep inelastic muon
  %proton scattering,''
  Phys.\ Lett.\  B {\bf 130} (1983) 118.
  %%CITATION = PHLTA,B130,118;%%

\bibitem{Arneodo:1986cf}
  M.~Arneodo {\it et al.}  [European Muon Collaboration],
  %``Measurement of hadron azimuthal distributions in deep inelastic muon proton
  %scattering,''
  Z.\ Phys.\  C {\bf 34} (1987) 277.
  %%CITATION = ZEPYA,C34,277;%%

\bibitem{Osipenko}
  M.~Osipenko {\it et al.}  [CLAS Collaboration],
  %``Measurement of semi-inclusive pi+ electroproduction off proton''
  Phys.\ Rev.\  D {\bf 80} (2009) 032004
  [arXiv:0809.1153 [hep-ex]].
  %%CITATION = PHRVA,D48,5057;%%

\bibitem{Mkrtchyan:2007sr}
  H.~Mkrtchyan {\it et al.},
  %``Transverse momentum dependence of semi-inclusive pion production,''
  Phys.\ Lett.\  B {\bf 665}, 20 (2008)
  [arXiv:0709.3020 [hep-ph]].
  %%CITATION = PHLTA,B665,20;%%

\bibitem{Airapetian:2009jy}
  A.~Airapetian {\it et al.}  [HERMES Collaboration],
  %``Transverse momentum broadening of hadrons produced in semi-inclusive
  %deep-inelastic scattering on nuclei,''
  Phys.\ Lett.\  B {\bf 684}, 114 (2010)
  [arXiv:0906.2478 [hep-ex]].
  %%CITATION = PHLTA,B684,114;%%

\bibitem{Adams:1993hs}
  M.~R.~Adams {\it et al.}  [E665 Collaboration],
  %``Perturbative QCD effects observed in 490-GeV deep inelastic muon
  %scattering,''
  Phys.\ Rev.\  D {\bf 48} (1993) 5057.
  %%CITATION = PHRVA,D48,5057;%%

\bibitem{Breitweg:2000qh}
  J.~Breitweg {\it et al.}  [ZEUS Collaboration],
  %``Measurement of azimuthal asymmetries in deep inelastic scattering,''
  Phys.\ Lett.\  B {\bf 481} (2000) 199
  [arXiv:hep-ex/0003017].
  %%CITATION = PHLTA,B481,199;%%

\bibitem{Chekanov:2006gt}
  S.~Chekanov {\it et al.}  [ZEUS Collaboration],
  %``Measurement of azimuthal asymmetries in neutral current deep inelastic
  %scattering at HERA,''
  Eur.\ Phys.\ J.\  C {\bf 51} (2007) 289
  [arXiv:hep-ex/0608053].
  %%CITATION = EPHJA,C51,289;%%

\bibitem{Airapetian:2002mf}
  A.~Airapetian {\it et al.}  [HERMES Collaboration],
  %``Measurement of single-spin azimuthal asymmetries in semi-inclusive
  %electroproduction of pions and kaons on a longitudinally polarised deuterium
  %target,''
  Phys.\ Lett.\  B {\bf 562} (2003) 182
  [arXiv:hep-ex/0212039].
  %%CITATION = PHLTA,B562,182;%%

\bibitem{Ageev:2006da}
  E.~S.~Ageev {\it et al.}  [COMPASS Collaboration],
  %``A new measurement of the Collins and Sivers asymmetries on a transversely
  %polarised deuteron target,''
  Nucl.\ Phys.\  B {\bf 765}, 31 (2007)
  [arXiv:hep-ex/0610068].
  %%CITATION = NUPHA,B765,31;%%

\bibitem{Avakian:2003pk}
  H.~Avakian {\it et al.}  [CLAS Collaboration],
  %``Measurement of beam-spin asymmetries for deep inelastic pi+
  %electroproduction,''
  Phys.\ Rev.\  D {\bf 69} (2004) 112004
  [arXiv:hep-ex/0301005].
  %%CITATION = PHRVA,D69,112004;%%



\bibitem{Airapetian:1999tv}
  A.~Airapetian {\it et al.}  [HERMES Collaboration],
  %``Observation of a single-spin azimuthal asymmetry in semi-inclusive pion
  %electro-production,''
  Phys.\ Rev.\ Lett.\  {\bf 84}, 4047 (2000)
  [arXiv:hep-ex/9910062],
  %%CITATION = HEP-EX 9910062;%%
%\bibitem{Avakian:1999rr}
%  H.~Avakian  [HERMES Collaboration],
%  %``Azimuthal single-spin asymmetries in semi-inclusive DIS from HERMES,''
%  Nucl.\ Phys.\ Proc.\ Suppl.\  {\bf 79}, 523 (1999).
%  %%CITATION = NUPHZ,79,523;%%
%\bibitem{Airapetian:2001eg}
  % A.~Airapetian {\it et al.}  [HERMES Collaboration],
  % ``Single-spin azimuthal asymmetries in electroproduction of neutral pions
  % in semi-inclusive deep-inelastic scattering,''
  Phys.\ Rev.\ D {\bf 64}, 097101 (2001)
  [arXiv:hep-ex/0104005],
  %%CITATION = HEP-EX 0104005;%%
%\bibitem{Airapetian:2002mf}
  %A.~Airapetian {\it et al.}  [HERMES Collaboration],
  % Measurement of single-spin azimuthal asymmetries in semi-inclusive
  % electroproduction of pions and kaons on a longitudinally polarised
  % deuterium target,
  Phys.\ Lett.\ B {\bf 562}, 182 (2003)
  [arXiv:hep-ex/0212039].
  %%CITATION = HEP-EX 0212039;%%

\bibitem{Airapetian:2005jc}
  A.~Airapetian {\it et al.} [HERMES Collaboration],
  %``Subleading-twist effects in single-spin asymmetries in semi-inclusive
  %deep-inelastic scattering on a longitudinally polarized hydrogen target,''
  Phys.\ Lett.\ B {\bf 622}, 14 (2005)
  [arXiv:hep-ex/0505042],
  %%CITATION = HEP-EX 0505042;%%
%\bibitem{Airapetian:2006rx}
  %A.~Airapetian {\it et al.}  [HERMES Collaboration],
  %``Beam-spin asymmetries in the azimuthal distribution of pion
  %electroproduction,''
  Phys.\ Lett.\  B {\bf 648} (2007) 164
  [arXiv:hep-ex/0612059].
  %%CITATION = PHLTA,B648,164;%%

\bibitem{Avakian:2005ps}
  H.~Avakian, P.~Bosted, V.~Burkert and L.~Elouadrhiri  [CLAS],
  %``New results on SIDIS SSA from Jefferson Lab,''
  AIP Conf.\ Proc.\  {\bf 792} (2005) 945
  [arXiv:nucl-ex/0509032].\\
  %%CITATION = NUCL-EX 0509032;%%
%\bibitem{Gohn:2009}
  W.~Gohn, H.~Avakian, K.~Joo, and M.~Ungaro,
  % "Beam spin asymmetries from semi-inclusive pion electroproduction 
  %in deep inelastic scattering",
  AIP Conf.\ Proc.\  {\bf 1149} (2009) 461.

\bibitem{Airapetian:2004tw}
  A.~Airapetian {\it et al.}  [HERMES Collaboration],
  %``Single-spin asymmetries in semi-inclusive deep-inelastic scattering on a
  %transversely polarized hydrogen target,''
  Phys.\ Rev.\ Lett.\  {\bf 94}, 012002 (2005)
  [arXiv:hep-ex/0408013].
  %%CITATION = HEP-EX 0408013;%%

\bibitem{Alexakhin:2005iw}
  V.~Y.~Alexakhin {\it et al.}  [COMPASS Collaboration],
  %``First measurement of the transverse spin asymmetries of the deuteron in
  %semi-inclusive deep inelastic scattering,''
  Phys.\ Rev.\ Lett.\  {\bf 94}, 202002 (2005)
  [arXiv:hep-ex/0503002].
  %%CITATION = HEP-EX 0503002;%%

\bibitem{Airapetian:2008sk}
  A.~Airapetian {\it et al.}  [HERMES Collaboration],
  %``Evidence for a Transverse Single-Spin Asymmetry in Leptoproduction of
  %pi+pi- Pairs,''
  JHEP {\bf 0806}, 017 (2008)
  [arXiv:0803.2367 [hep-ex]],
  %%CITATION = JHEPA,0806,017;%%
%\bibitem{Airapetian:2009ti}
  %A.~Airapetian {\it et al.}  [HERMES Collaboration],
  %``Observation of the Naive-T-odd Sivers Effect in Deep-Inelastic
  %Scattering,''
  Phys.\ Rev.\ Lett.\  {\bf 103}, 152002 (2009)
  [arXiv:0906.3918 [hep-ex]].
  %%CITATION = PRLTA,103,152002;%%

\bibitem{Giordano:2009hi}
  F.~Giordano and R.~Lamb  [HERMES Collaboration],
  %``Measurement of azimuthal asymmetries of the unpolarized cross section at
  %HERMES,''
  AIP Conf.\ Proc.\  {\bf 1149}, 423 (2009)
  [arXiv:0901.2438 [hep-ex]].\\
  %%CITATION = APCPC,1149,423;%%
%\bibitem{Lamb:2009zza}
  R.~Lamb and F.~Giordano  [HERMES Collaboration],
  %``Measurement of the cos(phi(h)) and cos(2phi(h)) azimuthal moments of the
  %deep inelastic scattering cross-section,''
  Nucl.\ Phys.\  A {\bf 827}, 225C (2009);
  %%CITATION = NUPHA,A827,225C;%%
%\bibitem{Lamb:2009zz}
  %R.~Lamb and F.~Giordano  [HERMES Collaboration],
  %``Recent measurements of the cos(n**phi(h)(azimuthal modulations of the
  %unpolarized deep inelastic scattering cross-section at HERMES,''
  AIP Conf.\ Proc.\  {\bf 1182}, 573 (2009).
  %%CITATION = APCPC,1182,573;%%

\bibitem{Kafer:2008ud}
  W.~K\"afer  [COMPASS Collaboration],
  %``Measurements of Unpolarized Azimuthal Asymmetries at COMPASS,''
  arXiv:0808.0114 [hep-ex].\\
  %%CITATION = ARXIV:0808.0114;%%
%\bibitem{Sbrizzai:2009fc}
  G.~Sbrizzai  [COMPASS Collaboration],
  %``Measurements of unpolarised azimuthal asymmetries at COMPASS,''
  arXiv:0902.0578 [hep-ex].\\
  %%CITATION = ARXIV:0902.0578;%%
%\bibitem{Bressan:2009eu}
  A.~Bressan  [COMPASS Collaboration],
  %``Azimuthal asymmetries in SIDIS off unpolarized targets at COMPASS,''
  arXiv:0907.5511 [hep-ex].\\
  %%CITATION = ARXIV:0907.5511;%%
%\bibitem{Schill:2009di}
  C.~Schill  [COMPASS collaboration],
  %``Transverse Spin Effects at COMPASS,''
  arXiv:0909.5287 [hep-ex].\\
  %%CITATION = ARXIV:0909.5287;%%
%\bibitem{Joosten:2009zz}
  R.~Joosten  [COMPASS Collaboration],
  %``Unpolarized azimuthal asymmetries from the COMPASS experiment at CERN,''
  AIP Conf.\ Proc.\  {\bf 1182}, 585 (2009).
  %%CITATION = APCPC,1182,585;%%

\bibitem{Ito:1981XX}
  A.~S.~Ito {\it et al.}  [FNAL-288]
  Phys.\ Rev.\  D {\bf 23} (1981) 604.

\bibitem{Badier:1981ti}
  J.~Badier {\it et al.}  [NA3 Collaboration],
  %``Angular Distributions In The Dimuon Hadronic Production At 150-Gev/C,''
  Z.\ Phys.\  C {\bf 11} (1981) 195.
  %%CITATION = ZEPYA,C11,195;%%

\bibitem{Palestini:1985zc}
  S.~Palestini {\it et al.},
  %``Pion Structure As Observed In The Reaction Pi- N $\to$ Mu+ Mu- X At
  %80-Gev/C,''
  Phys.\ Rev.\ Lett.\  {\bf 55} (1985) 2649.
  %%CITATION = PRLTA,55,2649;%%

\bibitem{Falciano:1986wk}
  S.~Falciano {\it et al.}  [NA10 Collaboration],
  %``Angular Distributions Of Muon Pairs Produced By 194-Gev/C Negative Pions,''
  Z.\ Phys.\  C {\bf 31} (1986) 513.
  %%CITATION = ZEPYA,C31,513;%%

\bibitem{Guanziroli:1987rp}
  M.~Guanziroli {\it et al.}  [NA10 Collaboration],
  %``ANGULAR DISTRIBUTIONS OF MUON PAIRS PRODUCED BY NEGATIVE PIONS ON DEUTERIUM
  %AND TUNGSTEN,''
  Z.\ Phys.\  C {\bf 37} (1988) 545.
  %%CITATION = ZEPYA,C37,545;%%

\bibitem{Conway:1989fs}
  J.~S.~Conway {\it et al.},
  %``Experimental Study Of Muon Pairs Produced By 252-Gev Pions On Tungsten,''
  Phys.\ Rev.\  D {\bf 39} (1989) 92.
  %%CITATION = PHRVA,D39,92;%%

\bibitem{Zhu:2006gx}
  L.~Y.~Zhu {\it et al.}  [FNAL-E866/NuSea Collaboration],
  %``Measurement of angular distributions of Drell-Yan dimuons in p + d
  %interaction at 800-GeV/c,''
  Phys.\ Rev.\ Lett.\  {\bf 99}, 082301 (2007)
  [arXiv:hep-ex/0609005].
  %%CITATION = PRLTA,99,082301;%%

\bibitem{Zhu:2008sj}
  L.~Y.~Zhu {\it et al.}  [FNAL E866/NuSea Collaboration],
  %``Measurement of Angular Distributions of Drell-Yan Dimuons in $p + p$
  %Interactions at 800 GeV/c,''
  Phys.\ Rev.\ Lett.\  {\bf 102}, 182001 (2009)
  [arXiv:0811.4589 [nucl-ex]].
  %%CITATION = PRLTA,102,182001;%%

\bibitem{Abe:2005zx}
  K.~Abe {\it et al.}  [Belle Collaboration],
  %``Measurement of azimuthal asymmetries in inclusive production of hadron
  %pairs in e+ e- annihilation at Belle,''
  Phys.\ Rev.\ Lett.\  {\bf 96}, 232002 (2006)
  [arXiv:hep-ex/0507063].\\
  %%CITATION = PRLTA,96,232002;%%
%\bibitem{Ogawa:2006bm}
  A.~Ogawa, M.~Grosse-Perdekamp, R.~Seidl and K.~Hasuko,
  %``Spin dependent fragmentation function at Belle,''
  arXiv:hep-ex/0607014.
  %%CITATION = HEP-EX/0607014;%%

\bibitem{Seidl:2008xc}
  R.~Seidl {\it et al.}  [Belle Collaboration],
  %``Measurement of Azimuthal Asymmetries in Inclusive Production of Hadron
  %Pairs in e+e- Annihilation at \sqrt{s} = 10.58 GeV,''
  Phys.\ Rev.\  D {\bf 78}, 032011 (2008)
  [arXiv:0805.2975 [hep-ex]].
  %%CITATION = PHRVA,D78,032011;%%

\bibitem{Vossen:2009xz}
  A.~Vossen, R.~Seidl, M.~Grosse-Perdekamp, M.~Leitgab, A.~Ogawa 
  and K.~Boyle,
  %``First Measurement of the Interference Fragmentation Function 
  % in $e^+e^-$ at Belle,''
  arXiv:0912.0353 [hep-ex].
  %%CITATION = ARXIV:0912.0353;%%

\bibitem{D'Alesio:2004up}
  U.~D'Alesio and F.~Murgia,
  %``Parton intrinsic motion in inclusive particle production: Unpolarized
  % cross sections, single spin asymmetries and the Sivers effect,''
  Phys.\ Rev.\ D {\bf 70}, 074009 (2004)
  [arXiv:hep-ph/0408092].
  %%CITATION = HEP-PH 0408092;%%

\bibitem{Anselmino:2005nn}
  M.~Anselmino, M.~Boglione, U.~D'Alesio, A.~Kotzinian, F.~Murgia and 
  A.~Prokudin,
  %``The role of Cahn and Sivers effects in deep inelastic scattering,''
  Phys.\ Rev.\  D {\bf 71} (2005) 074006
  [arXiv:hep-ph/0501196].
  %%CITATION = PHRVA,D71,074006;%%

\bibitem{Collins:2005ie}
  J.~C.~Collins, A.~V.~Efremov, K.~Goeke, S.~Menzel, A.~Metz and P.~Schweitzer,
  %``Sivers effect in semi-inclusive deeply inelastic scattering,''
  Phys.\ Rev.\  D {\bf 73} (2006) 014021
  [arXiv:hep-ph/0509076].
  %%CITATION = PHRVA,D73,014021;%%

\bibitem{Oganesian:1997jq}
  K.~A.~Oganesian, H.~R.~Avakian, N.~Bianchi and P.~Di Nezza,
  %``Investigations of azimuthal asymmetry in semi-inclusive  leptoproduction,''
  Eur.\ Phys.\ J.\  C {\bf 5}, 681 (1998)
  [arXiv:hep-ph/9709342].
  %%CITATION = EPHJA,C5,681;%%

\bibitem{Gamberg:2003ey}
  L.~P.~Gamberg, G.~R.~Goldstein and K.~A.~Oganessyan,
  %``Novel transversity properties in semi-inclusive deep inelastic
  %scattering,''
  Phys.\ Rev.\  D {\bf 67}, 071504 (2003)
  [arXiv:hep-ph/0301018].
  %%CITATION = PHRVA,D67,071504;%%

\bibitem{Barone:2005kt}
  V.~Barone, Z.~Lu and B.~Q.~Ma,
  %``On the cos(2phi) asymmetry in unpolarized leptoproduction,''
  Phys.\ Lett.\  B {\bf 632}, 277 (2006)
  [arXiv:hep-ph/0512145].
  %%CITATION = PHLTA,B632,277;%%

\bibitem{Barone:2006ws}
  V.~Barone, Z.~Lu and B.~Q.~Ma,
  %``The cos(2phi) asymmetry of Drell-Yan and J/psi production in unpolarized p
  %anti-p scattering,''
  Eur.\ Phys.\ J.\  C {\bf 49}, 967 (2007)
  [arXiv:hep-ph/0612350].
  %%CITATION = EPHJA,C49,967;%%

\bibitem{Gamberg:2007wm}
  L.~P.~Gamberg, G.~R.~Goldstein and M.~Schlegel,
  %``Transverse Quark Spin Effects and the Flavor Dependence of the Boer-Mulders
  %Function,''
  Phys.\ Rev.\  D {\bf 77}, 094016 (2008)
  [arXiv:0708.0324 [hep-ph]].
  %%CITATION = PHRVA,D77,094016;%%

\bibitem{Zhang:2008ez}
  B.~Zhang, Z.~Lu, B.~Q.~Ma and I.~Schmidt,
  %``$\cos 2 \phi$ asymmetries in unpolarized semi-inclusive DIS,''
  Phys.\ Rev.\  D {\bf 78}, 034035 (2008)
  [arXiv:0807.0503 [hep-ph]].
  %%CITATION = PHRVA,D78,034035;%%

\bibitem{Barone:2008tn}
  V.~Barone, A.~Prokudin and B.~Q.~Ma,
  %``A systematic phenomenological study of the $\cos 2 \phi$ asymmetry in
  %unpolarized semi--inclusive DIS,''
  Phys.\ Rev.\  D {\bf 78}, 045022 (2008)
  [arXiv:0804.3024 [hep-ph]].
  %%CITATION = PHRVA,D78,045022;%%

\bibitem{Barone:2009hw}
  V.~Barone, S.~Melis and A.~Prokudin,
  %``The Boer-Mulders effect in unpolarized SIDIS: an analysis of the COMPASS
  %and HERMES data on the $\cos 2 \phi$ asymmetry,''
  arXiv:0912.5194 [hep-ph].
  %%CITATION = ARXIV:0912.5194;%%

\bibitem{Efremov:2004tp}
  A.~V.~Efremov, K.~Goeke, S.~Menzel, A.~Metz and P.~Schweitzer,
  %``Sivers effect in semi-inclusive DIS and in the Drell-Yan process,''
  Phys.\ Lett.\  B {\bf 612}, 233 (2005)
  [arXiv:hep-ph/0412353].
  %%CITATION = PHLTA,B612,233;%%

\bibitem{Collins:2005rq}
  J.~C.~Collins {\it et al.},
  %``Sivers effect in Drell Yan at RHIC,''
  Phys.\ Rev.\  D {\bf 73}, 094023 (2006)
  [arXiv:hep-ph/0511272].
  %%CITATION = PHRVA,D73,094023;%%

\bibitem{Bianconi:2005yj}
  A.~Bianconi and M.~Radici,
  %``Monte Carlo simulation of the Sivers effect in high-energy proton  proton
  %collisions,''
  Phys.\ Rev.\  D {\bf 73}, 034018 (2006)
  [arXiv:hep-ph/0512091].
  %%CITATION = PHRVA,D73,034018;%%

\bibitem{Anselmino:2009st}
  M.~Anselmino, M.~Boglione, U.~D'Alesio, S.~Melis, F.~Murgia and A.~Prokudin,
  %``Sivers effect in Drell-Yan processes,''
  Phys.\ Rev.\  D {\bf 79}, 054010 (2009)
  [arXiv:0901.3078 [hep-ph]].
  %%CITATION = PHRVA,D79,054010;%%

\bibitem{Kang:2009sm}
  Z.~B.~Kang and J.~W.~Qiu,
  %``Single transverse spin asymmetry of dilepton production near $Z^0$ pole,''
  arXiv:0912.1319 [hep-ph].
  %%CITATION = ARXIV:0912.1319;%%

\bibitem{Bianconi:2006hc}
  A.~Bianconi and M.~Radici,
  %``Monte Carlo simulation of single spin asymmetries in pion proton
  %collisions,''
  Phys.\ Rev.\  D {\bf 73}, 114002 (2006)
  [arXiv:hep-ph/0602103].
  %%CITATION = PHRVA,D73,114002;%%


\bibitem{Sissakian:2008th}
  A.~Sissakian, O.~Shevchenko, A.~Nagaytsev and O.~Ivanov,
  %``Transversity, Boer-Mulders and Sivers distributions from Drell-Yan
  %processes with $pp$, $pD$ and $DD$ collisions,''
  Eur.\ Phys.\ J.\  C {\bf 59}, 659 (2009)
  [arXiv:0807.2480 [hep-ph]].
  %%CITATION = EPHJA,C59,659;%%

\bibitem{Sissakian:2010zz}
  A.~N.~Sissakian, O.~Y.~Shevchenko, A.~P.~Nagaitsev and O.~N.~Ivanov,
  %``Polarization effects in Drell-Yan processes,''
  Phys.\ Part.\ Nucl.\  {\bf 41}, 64 (2010).
  %%CITATION = PPNUE,41,64;%%

\bibitem{Bacchetta:2010si}
  A.~Bacchetta, M.~Radici, F.~Conti and M.~Guagnelli,
  %``Weighted azimuthal asymmetries in a diquark spectator model,''
  arXiv:1003.1328 [hep-ph].
  %%CITATION = ARXIV:1003.1328;%%

\bibitem{Gamberg:2005ip}
   L.~P.~Gamberg and G.~R.~Goldstein,
   %``T-odd effects in unpolarized Drell-Yan scattering,''
   Phys.\ Lett.\  B {\bf 650}, 362 (2007)
   [arXiv:hep-ph/0506127].
   %%CITATION = PHLTA,B650,362;%%

\bibitem{Sissakian:2005vd}
  A.~N.~Sissakian, O.~Y.~Shevchenko, A.~P.~Nagaytsev and O.~N.~Ivanov,
  %``An approach to direct extraction of transversity and its conjugate  T-odd
  %distribution from the unpolarized and single-polarized Drell-Yan
  %processes,''
  Phys.\ Rev.\  D {\bf 72}, 054027 (2005)
  [arXiv:hep-ph/0505214].
  %%CITATION = PHRVA,D72,054027;%%

\bibitem{Sissakian:2005yp}
  A.~Sissakian, O.~Shevchenko, A.~Nagaytsev, O.~Denisov and O.~Ivanov,
  %``Transversity and its accompanying T-odd distribution from Drell-Yan
  %processes with pion proton collisions,''
  Eur.\ Phys.\ J.\  C {\bf 46}, 147 (2006)
  [arXiv:hep-ph/0512095].
  %%CITATION = EPHJA,C46,147;%%

\bibitem{Lu:2009ip}
  Z.~Lu and I.~Schmidt,
  %``Updating Boer-Mulders functions from unpolarized pd and pp Drell-Yan
  %data,''
  Phys.\ Rev.\  D {\bf 81}, 034023 (2010) 
  [arXiv:0912.2031 [hep-ph]].
  %%CITATION = ARXIV:0912.2031;%%

\bibitem{Metz:2004je}
   A.~Metz and M.~Schlegel,
   %``Twist-3 single spin asymmetries in semi-inclusive deep-inelastic
   %scattering,''
   Eur.\ Phys.\ J.\  A {\bf 22}, 489 (2004)
   [arXiv:hep-ph/0403182].
   %%CITATION = EPHJA,A22,489;%%

\bibitem{Bacchetta:2004zf}
   A.~Bacchetta, P.~J.~Mulders and F.~Pijlman,
   %``New observables in longitudinal single-spin asymmetries in  semi-inclusive
   %DIS,''
   Phys.\ Lett.\  B {\bf 595}, 309 (2004)
   [arXiv:hep-ph/0405154].
   %%CITATION = PHLTA,B595,309;%%

\bibitem{Goeke:2005hb}
  K.~Goeke, A.~Metz and M.~Schlegel,
  %``Parameterization of the quark-quark correlator of a spin-1/2 hadron,''
  Phys.\ Lett.\ B {\bf 618}, 90 (2005)
  [arXiv:hep-ph/0504130].
  %%CITATION = HEP-PH 0504130;%%

\bibitem{Bacchetta:2006tn}
  A.~Bacchetta, M.~Diehl, K.~Goeke, A.~Metz, P.~J.~Mulders and M.~Schlegel,
  %``Semi-inclusive deep inelastic scattering at small transverse momentum,''
  JHEP {\bf 0702} (2007) 093
  [arXiv:hep-ph/0611265].
  %%CITATION = JHEPA,0702,093;%%

\bibitem{Hagiwara:1982cq}
  K.~Hagiwara, K.~I.~Hikasa and N.~Kai,
  %``Time Reversal Odd Asymmetry In Semiinclusive Leptoproduction In Quantum
  %Chromodynamics,''
  Phys.\ Rev.\  D {\bf 27} (1983) 84.
  %%CITATION = PHRVA,D27,84;%%

\bibitem{Gamberg:2006ru}
  L.~P.~Gamberg, D.~S.~Hwang, A.~Metz and M.~Schlegel,
  %``Light-cone divergence in twist-3 correlation functions,''
  Phys.\ Lett.\  B {\bf 639}, 508 (2006)
  [arXiv:hep-ph/0604022].
  %%CITATION = PHLTA,B639,508;%%

\bibitem{Bacchetta:2008xw}
   A.~Bacchetta, D.~Boer, M.~Diehl and P.~J.~Mulders,
   %``Matches and mismatches in the descriptions of semi-inclusive processes at
   %low and high transverse momentum,''
   JHEP {\bf 0808}, 023 (2008)
   [arXiv:0803.0227 [hep-ph]].
   %%CITATION = JHEPA,0808,023;%%

\bibitem{Avakian:2007mv}
  H.~Avakian, A.~V.~Efremov, K.~Goeke, A.~Metz, P.~Schweitzer and T.~Teckentrup,
  %``Are there approximate relations among transverse momentum dependent
  %distribution functions?,''
  Phys.\ Rev.\  D {\bf 77}, 014023 (2008)
  [arXiv:0709.3253 [hep-ph]].
  %%CITATION = PHRVA,D77,014023;%%

\bibitem{Metz:2008ib}
  A.~Metz, P.~Schweitzer and T.~Teckentrup,
  %``Lorentz invariance relations between parton distributions and the
  %Wandzura-Wilczek approximation,''
  Phys.\ Lett.\  B {\bf 680}, 141 (2009)
  [arXiv:0810.5212 [hep-ph]].
  %%CITATION = PHLTA,B680,141;%%

\bibitem{Teckentrup:2009tk}
  T.~Teckentrup, A.~Metz and P.~Schweitzer,
  %``Lorentz invariance relations and Wandzura-Wilczek approximation,''
  Mod.\ Phys.\ Lett.\  A {\bf 24}, 2950 (2009)
  [arXiv:0910.2567 [hep-ph]].
  %%CITATION = MPLAE,A24,2950;%%

\bibitem{Accardi:2009au}
  A.~Accardi, A.~Bacchetta, W.~Melnitchouk and M.~Schlegel,
  %``What can break the Wandzura--Wilczek relation?,''
  JHEP {\bf 0911}, 093 (2009)
  [arXiv:0907.2942 [hep-ph]].\\
  %%CITATION = JHEPA,0911,093;%%
%\bibitem{Accardi:2009nv}
  A.~Accardi, A.~Bacchetta and M.~Schlegel,
  %``What can we learn from the breaking of the Wandzura-Wilczek relation?,''
  AIP Conf.\ Proc.\  {\bf 1155}, 35 (2009)
  [arXiv:0905.3118 [hep-ph]].
  %%CITATION = APCPC,1155,35;%%

\bibitem{Avakian:2010br}
  H.~Avakian, A.~V.~Efremov, P.~Schweitzer and F.~Yuan,
  %``The transverse momentum dependent distribution functions in the bag
  %model,''
  arXiv:1001.5467 [hep-ph].
  %%CITATION = ARXIV:1001.5467;%%
%\bibitem{Avakian:2009jt}
  H.~Avakian, A.~V.~Efremov, P.~Schweitzer, O.~V.~Teryaev, F.~Yuan and P.~Zavada,
  %``Insights on non-perturbative aspects of TMDs from models,''
  Mod.\ Phys.\ Lett.\  A {\bf 24}, 2995 (2009)
  [arXiv:0910.3181 [hep-ph]].
  %%CITATION = MPLAE,A24,2995;%%

\bibitem{Schafer:1999kn}
  A.~Sch\"afer and O.~V.~Teryaev,
  %``Sum rules for the T-odd fragmentation functions,''
  Phys.\ Rev.\ D {\bf 61} (2000) 077903
  [arXiv:hep-ph/9908412].\\
  %%CITATION = HEP-PH 9908412;%%
%\bibitem{Meissner:2010cc}
  S.~Meissner, A.~Metz and D.~Pitonyak,
  %``Momentum sum rules for fragmentation functions,''
  arXiv:1002.4393 [hep-ph].
  %%CITATION = ARXIV:1002.4393;%%

\bibitem{Artru:1995bh}
  X.~Artru, J.~Czy\.zewski and H.~Yabuki,
  %``Single spin asymmetry in inclusive pion production, Collins effect and  the
  %string model,''
  Z.\ Phys.\  C {\bf 73} (1997) 527
  [arXiv:hep-ph/9508239].
  %%CITATION = ZEPYA,C73,527;%%

\bibitem{Vogelsang:2005cs}
  W.~Vogelsang and F.~Yuan,
  %``Single-transverse spin asymmetries: From DIS to hadronic collisions,''
  Phys.\ Rev.\  D {\bf 72} (2005) 054028
  [arXiv:hep-ph/0507266].
  %%CITATION = PHRVA,D72,054028;%%

\bibitem{Efremov:2006qm}
  A.~V.~Efremov, K.~Goeke and P.~Schweitzer,
  %``Collins effect in semi-inclusive deeply inelastic scattering and in e+  e-
  %annihilation,''
  Phys.\ Rev.\  D {\bf 73} (2006) 094025
  [arXiv:hep-ph/0603054].
  %%CITATION = PHRVA,D73,094025;%%

\bibitem{Anselmino:2007fs}
  M.~Anselmino, M.~Boglione, U.~D'Alesio, A.~Kotzinian, F.~Murgia, A.~Prokudin and C.~Turk,
  %``Transversity and Collins functions from SIDIS and e+ e- data,''
  Phys.\ Rev.\  D {\bf 75} (2007) 054032
  [arXiv:hep-ph/0701006].
  %%CITATION = PHRVA,D75,054032;%%

\bibitem{Anselmino:2008jk}
  M.~Anselmino, M.~Boglione, U.~D'Alesio, A.~Kotzinian, F.~Murgia, A.~Prokudin and S.~Melis,
  %``Update on transversity and Collins functions from SIDIS and e+ e- data,''
  Nucl.\ Phys.\ Proc.\ Suppl.\  {\bf 191}, 98 (2009)
  [arXiv:0812.4366 [hep-ph]].
  %%CITATION = NUPHZ,191,98;%%

\bibitem{Anselmino:2006rv}
  M.~Anselmino, M.~Boglione, A.~Prokudin and C.~Turk,
  %``Semi-inclusive deep inelastic scattering processes from small to large
  %P(T),''
  Eur.\ Phys.\ J.\  A {\bf 31}, 373 (2007)
  [arXiv:hep-ph/0606286].
  %%CITATION = EPHJA,A31,373;%%

\bibitem{Drell:1970wh}
  S.~D.~Drell and T.~M.~Yan,
  %``Massive Lepton Pair Production In Hadron-Hadron Collisions At
  %High-Energies,''
  Phys.\ Rev.\ Lett.\  {\bf 25} (1970) 316
  [Erratum-ibid.\  {\bf 25} (1970) 902];
  %%CITATION = PRLTA,25,316;%%
%\bibitem{Drell:1970yt}
  % S.~D.~Drell and T.~M.~Yan,
  %``Partons and their applications at high energies,''
  Annals Phys.\  {\bf 66} (1971) 578.
  %[Annals Phys.\  {\bf 281} (2000) 450].
  %%CITATION = APNYA,281,450;%%

\bibitem{Christenson:1970um}
  J.~H.~Christenson, G.~S.~Hicks, L.~M.~Lederman, P.~J.~Limon, B.~G.~Pope, % and 
  E.~Zavattini,
  %``Observation Of Massive Muon Pairs In Hadron Collisions,''
  Phys.\ Rev.\ Lett.\  {\bf 25} (1970) 1523.
  %%CITATION = PRLTA,25,1523;%%

\bibitem{Stirling:1993gc}
  W.~J.~Stirling and M.~R.~Whalley,
  %``A Compilation of Drell-Yan cross-sections,''
  J.\ Phys.\ G {\bf 19} (1993) D1.
  %%CITATION = JPHGB,G19,D1;%%

\bibitem{McGaughey:1999mq}
  P.~L.~McGaughey, J.~M.~Moss and J.~C.~Peng,
  %``High-energy hadron-induced dilepton production from nucleons and  nuclei,''
  Ann.\ Rev.\ Nucl.\ Part.\ Sci.\  {\bf 49} (1999) 217
  [arXiv:hep-ph/9905409].
  %%CITATION = ARNUA,49,217;%%

\bibitem{Reimer:2007iy}
  P.~E.~Reimer,
  %``Exploring the partonic structure of hadrons through the Drell-Yan
  %process,''
  J.\ Phys.\ G {\bf 34}, S107 (2007)
  [arXiv:0704.3621 [nucl-ex]].
  %%CITATION = JPHGB,G34,S107;%%

\bibitem{Tangerman:1994eh}
   R.~D.~Tangerman and P.~J.~Mulders,
   %``Intrinsic transverse momentum and the polarized Drell-Yan process,''
   Phys.\ Rev.\  D {\bf 51}, 3357 (1995)
   [arXiv:hep-ph/9403227].
   %%CITATION = PHRVA,D51,3357;%%

\bibitem{Arnold:2008kf}
  S.~Arnold, A.~Metz and M.~Schlegel,
  %``Dilepton production from polarized hadron hadron collisions,''
  Phys.\ Rev.\  D {\bf 79}, 034005 (2009)
  [arXiv:0809.2262 [hep-ph]].
  %%CITATION = PHRVA,D79,034005;%%

\bibitem{Vogelsang:1992jn}
  W.~Vogelsang and A.~Weber,
  %``Drell-Yan dimuon production with transversely polarized protons,''
  Phys.\ Rev.\  D {\bf 48} (1993) 2073.
  %%CITATION = PHRVA,D48,2073;%%

\bibitem{Contogouris:ws}
    A.~P.~Contogouris, B.~Kamal and Z.~Merebashvili,
    %``One Loop Corrections To Lepton Pair Production
    % By Transversely Polarized Hadrons,''
    Phys.\ Lett.\ B {\bf 337} (1994) 169. %%CITATION = PHLTA,B337,169;%%

\bibitem{Ratcliffe:2004we}
  P.~G.~Ratcliffe,
  %``Transversity K factors for Drell-Yan,''
  Eur.\ Phys.\ J.\  C {\bf 41} (2005) 319
  [arXiv:hep-ph/0412157].
  %%CITATION = EPHJA,C41,319;%%

\bibitem{Shimizu:2005fp}
  H.~Shimizu, G.~Sterman, W.~Vogelsang and H.~Yokoya,
  %``Dilepton production near partonic threshold in transversely polarized
  %proton-antiproton collisions,''
  Phys.\ Rev.\  D {\bf 71} (2005) 114007
  [arXiv:hep-ph/0503270].
  %%CITATION = PHRVA,D71,114007;%%

\bibitem{Barone:2005cr}
  V.~Barone, A.~Cafarella, C.~Coriano', M.~Guzzi and P.~Ratcliffe,
  %``Double transverse-spin asymmetries in Drell-Yan processes with
  %antiprotons,''
  Phys.\ Lett.\  B {\bf 639} (2006) 483
  [arXiv:hep-ph/0512121].
  %%CITATION = PHLTA,B639,483;%%

\bibitem{Kawamura:2007ze}
  H.~Kawamura, J.~Kodaira and K.~Tanaka,
  %``Soft gluon corrections to double transverse-spin asymmetries for
  %small-$Q_T$ dilepton production at RHIC and J-PARC,''
  Nucl.\ Phys.\  B {\bf 777} (2007) 203
  [arXiv:hep-ph/0703079].
  %%CITATION = NUPHA,B777,203;%%


\bibitem{Berger:1979du}
  E.~L.~Berger and S.~J.~Brodsky,
  %``Quark Structure Functions Of Mesons And The Drell-Yan Process,''
  Phys.\ Rev.\ Lett.\  {\bf 42} (1979) 940.\\
  %%CITATION = PRLTA,42,940;%%
%\bibitem{Berger:1979xz}
  E.~L.~Berger,
  %``Quark Structure Functions Of Mesons, Fragmentation Functions, Higher Twist
  %Effects In QCD, Deep Inelastic Scattering, And The Drell-Yan Process,''
  Z.\ Phys.\  C {\bf 4} (1980) 289.
  %%CITATION = ZEPYA,C4,289;%%

\bibitem{Badier:1982zb}
  J.~Badier {\it et al.}  [NA3 Collaboration],
  %``Measurement Of The Transverse Momentum Of Dimuons Produced By Hadronic
  %Interactions At 150-Gev/C, 200-Gev/C And 280-Gev/C,''
  Phys.\ Lett.\  B {\bf 117} (1982) 372.
  %%CITATION = PHLTA,B117,372;%%

\bibitem{Bordalo:1987cs}
  P.~Bordalo {\it et al.}  [NA10 Collaboration],
  %``NUCLEAR EFFECTS ON THE NUCLEON STRUCTURE FUNCTIONS IN HADRONIC HIGH MASS
  %DIMUON PRODUCTION,''
  Phys.\ Lett.\  B {\bf 193} (1987) 368;
  %%CITATION = PHLTA,B193,368;%%
%\bibitem{Bordalo:1987cr}
  %P.~Bordalo {\it et al.}  [NA10 Collaboration],
  %``OBSERVATION OF A NUCLEAR DEPENDENCE OF THE TRANSVERSE MOMENTUM DISTRIBUTION
  %OF MASSIVE MUON PAIRS PRODUCED IN HADRONIC COLLISIONS,''
  Phys.\ Lett.\  B {\bf 193} (1987) 373.
  %%CITATION = PHLTA,B193,373;%%

\bibitem{Collins:1977iv}
  J.~C.~Collins and D.~E.~Soper,
  %``Angular Distribution Of Dileptons In High-Energy Hadron Collisions,''
  Phys.\ Rev.\  D {\bf 16}, 2219 (1977).
  %%CITATION = PHRVA,D16,2219;%%

\bibitem{Lam:1978pu}
  C.~S.~Lam and W.~K.~Tung,
  %``A Systematic Approach To Inclusive Lepton Pair Production In Hadronic
  %Collisions,''
  Phys.\ Rev.\  D {\bf 18} (1978) 2447;
  %%CITATION = PHRVA,D18,2447;%% 
%\bibitem{Lam:1980uc}
  % C.~S.~Lam and W.~K.~Tung,
  %``A Parton Model Relation Sans QCD Modifications In Lepton Pair
  %Productions,''
  Phys.\ Rev.\  D {\bf 21} (1980) 2712.
  %%CITATION = PHRVA,D21,2712;%%

\bibitem{Collins:1978yt}
  J.~C.~Collins,
  %``Simple Prediction Of QCD For Angular Distribution Of Dileptons In Hadron
  %Collisions,''
  Phys.\ Rev.\ Lett.\  {\bf 42}, 291 (1979).
  %%CITATION = PRLTA,42,291;%%

\bibitem{Mirkes:1994dp}
  E.~Mirkes and J.~Ohnemus,
  %``Angular distributions of Drell-Yan lepton pairs at the Tevatron: Order
  %$\alpha-s^{2}$ corrections and Monte Carlo studies,''
  Phys.\ Rev.\  D {\bf 51}, 4891 (1995)
  [arXiv:hep-ph/9412289].
  %%CITATION = PHRVA,D51,4891;%%

\bibitem{Zhou:2009rp}
  J.~Zhou, F.~Yuan and Z.~T.~Liang,
  %``Drell-Yan Lepton Pair Azimuthal Asymmetry in Hadronic Processes,''
  Phys.\ Lett.\  B {\bf 678}, 264 (2009)
  [arXiv:0901.3601 [hep-ph]].
  %%CITATION = PHLTA,B678,264;%%

\bibitem{Malhotra:1982cx}
  P.~K.~Malhotra,
%  %``Transverse Momentum Spectra Of Dimuons Produced In Hadronic Interactions
%  %And Comparison With QCD,'' 
in Proceedings of
 ``Drell Yan Workshop'', Batavia, Ill., Oct 7-8, 1982 [C82/10/07].\\
%  %%CITATION = PRINT-82-0846-FERMILAB-;%%
%\bibitem{Cox:1982wy}
  B.~Cox and P.~K.~Malhotra,
  %``Comparison Of Energy Dependence Of Transverse Momentum Of Dimuons Produced
  %In P N And Pi- N Interactions With QCD Predictions,''
  Phys.\ Rev.\  D {\bf 29} (1984) 63.
  %%CITATION = PHRVA,D29,63;%%

\bibitem{Landry:2002ix}
  F.~Landry, R.~Brock, P.~M.~Nadolsky and C.~P.~Yuan,
  %``Tevatron Run-1 Z boson data and Collins-Soper-Sterman resummation
  %formalism,''
  Phys.\ Rev.\  D {\bf 67} (2003) 073016
  [arXiv:hep-ph/0212159].
  %%CITATION = PHRVA,D67,073016;%%

\bibitem{Idilbi:2004vb}
  A.~Idilbi, X.~d.~Ji, J.~P.~Ma and F.~Yuan,
  %``Collins-Soper equation for the energy evolution of transverse-momentum  and
  %spin dependent parton distributions,''
  Phys.\ Rev.\  D {\bf 70}, 074021 (2004)
  [arXiv:hep-ph/0406302].
  %%CITATION = PHRVA,D70,074021;%%

\bibitem{Cherednikov:2007tw}
  I.~O.~Cherednikov and N.~G.~Stefanis,
  %``Renormalization, Wilson lines, and transverse-momentum dependent parton
  %distribution functions,''
  Phys.\ Rev.\  D {\bf 77}, 094001 (2008)
  [arXiv:0710.1955 [hep-ph]].
  %%CITATION = PHRVA,D77,094001;%%
%\bibitem{Cherednikov:2008ua}
  %I.~O.~Cherednikov and N.~G.~Stefanis,
  %``Wilson lines and transverse-momentum dependent parton distribution
  %functions: A renormalization-group analysis,''
  Nucl.\ Phys.\  B {\bf 802}, 146 (2008)
  [arXiv:0802.2821 [hep-ph]].
  %%CITATION = NUPHA,B802,146;%%
%\bibitem{Cherednikov:2009wk}
  %I.~O.~Cherednikov and N.~G.~Stefanis,
  %``Renormalization-group properties of transverse-momentum dependent parton
  %distribution functions in the light-cone gauge with the Mandelstam-Leibbrandt
  %prescription,''
  Phys.\ Rev.\  D {\bf 80}, 054008 (2009)
  [arXiv:0904.2727 [hep-ph]];
%\bibitem{Stefanis:2009ij}
  %N.~G.~Stefanis and I.~O.~Cherednikov,
  %``Renormalization-group anatomy of transverse-momentum dependent parton
  %distribution functions in QCD,''
  Mod.\ Phys.\ Lett.\  A {\bf 24}, 2913 (2009)
  [arXiv:0910.3108 [hep-ph]].
  %%CITATION = MPLAE,A24,2913;%%

\bibitem{Boffi:2009sh}
  S.~Boffi, A.~V.~Efremov, B.~Pasquini and P.~Schweitzer,
  %``Azimuthal spin asymmetries in light-cone constituent quark models,''
  Phys.\ Rev.\  D {\bf 79}, 094012 (2009)
  [arXiv:0903.1271 [hep-ph]].
  %%CITATION = PHRVA,D79,094012;%%

\bibitem{lattice-TMD}
  Ph.~H\"agler, B.~U.~Musch, J.~W.~Negele, A.~Sch\"afer,
  Europhys.~Lett.~88, 61001 (2009)
  [arXiv:0908.1283v1 [hep-lat]].\\
%\bibitem{Musch:2008jd}
  B.~U.~Musch, P.~H\"agler, A.~Sch\"afer, D.~B.~Renner, J.~W.~Negele
  [LHPC Collaboration],
  %``Transverse momentum distributions of quarks in the nucleon from Lattice
  %QCD,''
  PoS {\bf LC2008}, 053 (2008)
  [arXiv:0811.1536 [hep-lat]].
  %%CITATION = POSCI,LC2008,053;%%




\end{thebibliography}
\end{document}